\documentclass[utf8]{FrontiersinHarvard} 
\usepackage{url,hyperref,lineno,microtype,subcaption}
\usepackage[onehalfspacing]{setspace}

\def\keyFont{\fontsize{8}{11}\helveticabold }
\def\firstAuthorLast{Das {et~al.}}
\def\Authors{Susmita Das\,$^{1,2,3*}$, Anupam Bhardwaj\,$^{1}$ and Marcella Marconi\,$^{4}$}
\def\Address{$^{1}$Inter-University Center for Astronomy and Astrophysics (IUCAA), Post Bag 4, Ganeshkhind, Pune 411007, India \\
$^{2}$Konkoly Observatory, Research Centre for Astronomy and Earth Sciences, HUN-REN, Konkoly-Thege Mikl\'os \'ut 15-17, H-1121, Budapest, Hungary\\
$^{3}$CSFK, MTA Centre of Excellence, Budapest, Konkoly Thege Miklós út 15-17., H-1121, Hungary\\
$^{4}$INAF-Osservatorio Astronomico di Capodimonte, Salita Moiariello 16, 80131, Naples, Italy}
\def\corrAuthor{Susmita Das}

\def\corrEmail{susmita.das@iucaa.in}

\begin{document}
\onecolumn
\firstpage{1}

\title[Type II Cepheids]{Type~II Cepheids: Period-Luminosity-Metallicity relations for the Population~II distance scale} 

\author[\firstAuthorLast ]{\Authors}
\address{}
\correspondance{}

\extraAuth{}

\maketitle

\begin{abstract}

Type~II Cepheids are a class of pulsating variable stars that play a critical role in our understanding of stellar evolution, distance measurement and tracing the structure and kinematics of old stars in nearby galaxies. This review provides a comprehensive summary of the current state of research on Type~II Cepheids, including their observed properties, pulsation mechanisms and their distinction from other variable stars. These pulsating variable stars, found primarily in older stellar populations, exhibit well-defined period-luminosity relations but with an added advantage that they exhibit weak or negligible dependence on metallicity. We explore their relevance in the context of their role as distance indicators and potential calibrators of the first rung of the extragalactic distance ladder.  Finally, the review highlights recent advancements in theoretical models, observations across different wavelengths and ongoing debates concerning their classification.

\tiny
 \keyFont{ \section{Keywords:} classical pulsators, cepheid variables, population~II stars,  low-mass stars, period-luminosity relations, metallicity, evolution, pulsation} 
\end{abstract}

\section{Introduction}

Cepheid variables have long served as vital distance indicators due to their well-defined period-luminosity ($PL$) relations \citep[see,][among others]{soszynski2011, soszynski2017}. Among these, Type~II Cepheids (T2CEPs) belong to the older, low-mass class of classical pulsators, distinct from their more massive and younger classical (Type~I) counterparts \citep[see reviews,][]{gingold1985, wallerstein2002, bhardwaj2020, bono2024}. T2CEPs are predominantly found in old stellar populations $-$ such as the Galactic halo and bulge, globular clusters and the Magellanic Clouds $-$ as reported by the OGLE surveys \citep{soszynski2017, soszynski2018} and the Gaia catalogs \citep{ripepi2023}, and have also been identified in the Andromeda group \citep{kodric2018}. At a given pulsation period, T2CEPs are 1-2 magnitudes fainter than classical Cepheids. However, their tight $PL$ relations make them extremely important for anchoring the Population~II distance scale, especially when used in combination with RR~Lyraes and the Tip of the Red Giant Branch (TRGB) \citep[see,][]{majaess2010, braga2020, das2025b, lengen2025}. They are particularly important astrophysical objects in metal-poor systems where classical Cepheids are scarce or absent \citep[see reviews,][]{caputo1998, wallerstein2002, sandage2006} and in globular clusters with very blue horizontal branch morphology with rare or no RR~Lyrae stars \citep{gingold1976, pritzl2002, dicriscienzo2007}.

The role of metallicity in calibrating the $PL$ relations of classical pulsators has gained increasing attention, especially with the advent of high-precision astrometry (e.g., from Gaia\footnote{\url{https://gea.esac.esa.int/archive/}}), photometric surveys (e.g., OGLE\footnote{\url{https://ogle.astrouw.edu.pl/}}, VISTA\footnote{\url{https://www.eso.org/sci/facilities/paranal/telescopes/vista.html}}), and spectroscopic datasets (e.g., APOGEE\footnote{\url{https://www.sdss4.org/dr17/irspec/}}, LAMOST\footnote{\url{https://www.lamost.org}}). Classical Cepheids are known to exhibit strong effects of metallicity on their $PL$ relations, especially at shorter wavelengths \citep[see, for example,][]{desomma2022, bhardwaj2024, ripepi2025, breuval2025}. On the other hand, even with a wide range of metallicities from [Fe/H] = –2.4 to –0.5 \citep[see Appendix~A,][]{bono2020}, T2CEPs seem to exhibit little or no metallicity dependence in their $PL$ relations, even in the optical bands (further details in Section~\ref{sec:PLZ}).

In this review, we sum up the current theoretical and empirical understanding of the metallicity effects on T2CEP $PL$ relations. We incorporate findings from key observational datasets as well as theoretical pulsation models and thereby explore the behaviour of the period-luminosity-metallcity ($PLZ$) relations across the different T2CEP subclasses (BL~Her, W~Vir, RV~Tau) and their implications for cosmic distance scaling.

\section{Period-Based Classification and Evolution of T2CEPs}

Based on their pulsation periods and evolutionary states, T2CEPs are sub-divided into three main classes following the classification suggested by \citet{soszynski2018}: the BL~Herculis stars (BL~Her), the W~Virginis stars (W~Vir) and the RV~Tauri stars (RV~Tau). In addition, there is a fourth sub-class $-$ the peculiar W~Vir stars (pW~Vir) that are bluer and brighter than W~Vir \citep{soszynski2008}. Understanding these subclasses is crucial because their evolutionary origins, pulsation characteristics and metallicity distributions differ, potentially influencing their $PL$ behaviour and the extent to which metallicity affects their luminosities.

BL~Her stars typically have short pulsation periods with $1 < P (\rm days) < 4$. They are believed to be low-mass stars evolving from the horizontal branch toward the asymptotic giant branch (AGB) \citep{bono2020}. A pulsation period of 1 day has traditionally been used to distinguish RR~Lyrae stars from T2CEPs \citep{soszynski2008, soszynski2014}; however, defining the boundary between RR~Lyraes and T2CEPs has long been problematic. Recent studies recommend using a more general, evolution-dependent criterion for this separation \citep{braga2020, bono2024}. W~Vir stars have intermediate periods with $4 < P (\rm days) < 20$ and are thought to be stars undergoing blue loops on the AGB or post-AGB evolutionary stages \citep{bono2020}. They often show more complex light curves and some evidence of binarity \citep{groenewegen2017a}. This is particularly evident in pW~Vir stars $-$ about 50\% of those identified in the Magellanic Clouds display signatures of binarity \citep{groenewegen2017a, soszynski2018}. RV Tau stars are long-period with $P (\rm days) > 20$, evolved post-AGB objects \citep{bono2020}. They exhibit alternating deep and shallow minima in their light curves, suggesting complex pulsation modes, and are frequently associated with circumstellar dust and mass loss. 

\section{The effect of metallicity on $PL$ relations}
\label{sec:PLZ}

Similar to classical Cepheids, T2CEPs obey well-defined $PL$ relations which make them important astrophysical objects, especially for old stellar systems. However, prior to the identification of Populations I and II by \citet{baade1944}, it was assumed that a single $PL$ relation was applicable to all Cepheids, leading to discrepancies in Hubble's early distance scale measurements \citep[][and references therein]{hubble1931} by a factor of two. It was \citet{baade1956} who concluded that ``there was no a priori reason to expect that two Cepheids of the same period, the one a member of Population I, the other of Population II, should have the same luminosity''\footnote{The explanation for the different $PL$ relations for type I and type II Cepheids is as follows: two Cepheids with equal mean radius and mean effective temperature would have the same luminosities (Stefan-Boltzmann law). If these two Cepheids obey Ritter’s pulsation equation $P \sqrt{<\rho>} = Q$ such that they have similar pulsation constants $Q$ and assuming typical masses for type I and type II Cepheids, we find that the expected pulsation period of a 0.6M type II Cepheid is three times longer than that of a 6M type I Cepheid \citep{catelan2015}.}, thereby subsequently resulting in the revised cosmic distance scale. 

Since then, there have been quite a few studies dedicated towards studying the different aspects of T2CEPS, on both the theoretical \citep{carson1982, kovacs1988, bono1997,deka2024} and the empirical \citep{gonzalez1994, bersier1997, balog1997, vinko1998, kiss2007, wielgorski2024, yacob2025} fronts. For the most recent empirical $PL$ relations of T2CEPs in multiple wavelengths, the interested reader is referred to the works of \citet{sicignano2024}, \citet{cruzreyes2025} and \citet{narloch2025}. However, in this review, our primary focus will be on studies that examine the role of metallicity on the $PL$ relations of T2CEPs.

One of the earliest $PLZ$ studies for T2CEPs was carried out by \citet{nemec1994}, with the assumed metal abundance [Fe/H] determined generally on the scale of \citet{zinn1984}. However, the paper follows an old classification with $P (d)<10$ as BL~Her, $10<P (d)<26$ as W~Vir stars and $26<P (d)<100$ as RV~Tau stars as was used in \citet{joy1949, demers1974, arp1955, wallerstein1984}. Therefore, while the combined $PLZ$ relations remain the same, the relations from individual sub-classes may be different. \citet{mcnamara1995} followed up the work of \citet{nemec1994} using the same observed dataset and obtained different $PL$ relations for the two classes of BL~Her and W~Vir stars when considered separately and was one of the firsts to rule out the dependence of the absolute magnitudes of the variables on [Fe/H]. They also ruled out the possibility of T2CEPS pulsating in the first-overtone mode, as was earlier suggested by \citet{arp1955} and \citet{nemec1994}. Note that while this broadly remains true, two first-overtone T2CEPs were recently identified in the Large Magellanic Cloud (LMC) by \citet{soszynski2019}. Although not a study that explores $PLZ$, \citet{alcock1998} as part of the MACHO project is interesting because unlike previous studies which explicitly excluded RV Tau stars from T2CEP $PL$ relations, they observed a single period-luminosity-color relation for both T2CEPs and RV Tau stars ($0.9 < \log(P) < 1.75$) in the LMC, thereby suggesting a common evolutionary channel.

\begin{table*}
\caption{Compilation of all $PLZ$ relations of the mathematical form, $M_\lambda=\alpha+\beta\log(P)+\gamma\mathrm{[Fe/H]}$ for T2CEPs available in the literature at different wavelengths.}
\centering
\scalebox{0.64}{
\begin{tabular}{c c c c c c c c}
\hline
Band & $\alpha$ & $\beta$ & $\gamma$ & $\sigma$ & $N$ & Data & Source\\
\hline
\multicolumn{8}{c}{BL Her}\\
\hline
Bol  & -0.131 $\pm$ 0.012 & -1.83 $\pm$ 0.029 & 0.047 $\pm$ 0.006 & 0.251 & 3266 & Theoretical & \citet{das2021}\\
$U$ & 0.438 $\pm$ 0.017 & -0.998 $\pm$ 0.041 & 0.234 $\pm$ 0.009 & 0.357 & 3266 & Theoretical & \citet{das2021}\\
$B$ & 0.276 $\pm$ 0.016 & -1.221 $\pm$ 0.04 & 0.082 $\pm$ 0.009 & 0.347 & 3266 & Theoretical & \citet{das2021}\\
$V$ & -0.134 $\pm$ 0.013 & -1.62 $\pm$ 0.032 & 0.006 $\pm$ 0.007 & 0.284 & 3266 & Theoretical & \citet{das2021}\\
$R$ & -0.372 $\pm$ 0.012 & -1.848 $\pm$ 0.028 & -0.007 $\pm$ 0.006 & 0.248 & 3266 & Theoretical & \citet{das2021}\\
$I$ & -0.593 $\pm$ 0.01 & -2.043 $\pm$ 0.025 & -0.001 $\pm$ 0.006 & 0.219 & 3266 & Theoretical & \citet{das2021}\\
$J$  & -0.28 $\pm$ 0.35 & -3.19 $\pm$ 0.38 & 0.2 $\pm$ 0.17 & 0.08 & 7 & Empirical & \citet{matsunaga2006}\\
$J$ & -0.908 $\pm$ 0.009 & -2.306 $\pm$ 0.021 & 0.005 $\pm$ 0.005 & 0.186 & 3266 & Theoretical & \citet{das2021}\\
$J$ & $-$ & $-$ & -0.387$\pm$0.123 & $-$ & 5 & Empirical & \citet{wielgorski2022}\\
$H$& -0.56 $\pm$ 0.38 & -3.25 $\pm$ 0.41 & 0.2 $\pm$ 0.19 & 0.09 & 7 & Empirical & \citet{matsunaga2006}\\
$H$ & -1.154 $\pm$ 0.007 & -2.58 $\pm$ 0.018 & 0.015 $\pm$ 0.004 & 0.156 & 3266 & Theoretical & \citet{das2021}\\
$H$ & $-$ & $-$ & -0.186$\pm$0.090 & $-$ & 5 & Empirical & \citet{wielgorski2022}\\
$K_S$& -0.71 $\pm$ 0.34 & -3.07 $\pm$ 0.37 & 0.15 $\pm$ 0.17 & 0.08 & 7& Empirical & \citet{matsunaga2006}\\
$K$ & -1.107 $\pm$ 0.008 & -2.539 $\pm$ 0.018 & 0.015 $\pm$ 0.004 & 0.16 & 3266 & Theoretical & \citet{das2021}\\
$K_S$ & $-$ & $-$ & -0.203$\pm$0.111 & $-$ & 5 & Empirical & \citet{wielgorski2022}\\
$L$ & -1.212 $\pm$ 0.007 & -2.611 $\pm$ 0.018 & 0.017 $\pm$ 0.004 & 0.154 & 3266 & Theoretical & \citet{das2021}\\
$L'$ & -1.214 $\pm$ 0.007 & -2.611 $\pm$ 0.018 & 0.019 $\pm$ 0.004 & 0.155 & 3266 & Theoretical & \citet{das2021}\\
$M$ & -1.441 $\pm$ 0.007 & -2.833 $\pm$ 0.016 & 0.047 $\pm$ 0.004 & 0.14 & 3266 & Theoretical & \citet{das2021}\\
$G$ & -0.262 $\pm$ 0.012 & -1.761 $\pm$ 0.03 & 0.001 $\pm$ 0.007 & 0.262 & 3266 & Theoretical & \citet{das2024}\\
$G_{BP}$ & -0.058 $\pm$ 0.014 & -1.555 $\pm$ 0.033 & 0.018 $\pm$ 0.008 & 0.292 & 3266 & Theoretical & \citet{das2024}\\
$G_{RP}$ & -0.601 $\pm$ 0.011 & -2.007 $\pm$ 0.026 & -0.012 $\pm$ 0.006 & 0.225 & 3266 & Theoretical & \citet{das2024}\\
$W_{VI}$ & -1.304 $\pm$ 0.007 & -2.698 $\pm$ 0.017 & -0.011 $\pm$ 0.004 & 0.146 & 3266 & Theoretical & \citet{das2021}\\
$W_{JK_S}$ & -1.0 $\pm$ 0.35 & -2.98 $\pm$ 0.38 & 0.12 $\pm$ 0.17 & 0.08 & 7& Empirical & \citet{matsunaga2006}\\
$W_{JK}$ & -1.244 $\pm$ 0.007 & -2.699 $\pm$ 0.017 & 0.023 $\pm$ 0.004 & 0.147 & 3266 & Theoretical & \citet{das2021}\\
$W_{JK_S}$ & $-$ & $-$ & -0.107$\pm$0.101 & $-$ & 5 & Empirical & \citet{wielgorski2022}\\
$W_{G_{BP}, G_{RP}}^G$ & -1.294 $\pm$ 0.007 & -2.619 $\pm$ 0.018 & -0.056 $\pm$ 0.004 & 0.154 & 3266 & Theoretical & \citet{das2024}\\
$W_{gi \rm{LSST}}$ & -0.501 $\pm$ 0.007 & -2.722 $\pm$ 0.017 & -0.065 $\pm$ 0.004 & 0.146 & 3266 & Theoretical & \citet{das2025b}\\
$W_{iz\rm{LSST}}$ & -0.123 $\pm$ 0.009 & -2.4 $\pm$ 0.021 & 0.024 $\pm$ 0.005 & 0.18 & 3266 & Theoretical & \citet{das2025b}\\
$W_{gy\rm{LSST}}$ & -0.306 $\pm$ 0.008 & -2.54 $\pm$ 0.019 & -0.026 $\pm$ 0.004 & 0.163 & 3266 & Theoretical & \citet{das2025b}\\
\hline
\multicolumn{8}{c}{W Vir}\\
\hline
$J$ & -1.19 $\pm$ 0.21 & -2.23 $\pm$ 0.15 & -0.22 $\pm$ 0.09 & 0.14 & 26 & Empirical & \citet{matsunaga2006}\\
$H$& -1.63 $\pm$ 0.21 & -2.24 $\pm$ 0.15 & -0.26 $\pm$ 0.09 & 0.14 & 26 & Empirical & \citet{matsunaga2006}\\
$K_S$ & -1.47 $\pm$ 0.22 & -2.3 $\pm$ 0.16 & -0.17 $\pm$ 0.09 & 0.14 & 26& Empirical & \citet{matsunaga2006}\\
$W_{JK_S}$ & -1.67 $\pm$ 0.25 & -2.35 $\pm$ 0.18 & -0.14 $\pm$ 0.11 & 0.16 & 26& Empirical & \citet{matsunaga2006}\\

\hline
\multicolumn{8}{c}{RV Tau}\\
\hline
$J$ & -1.39 $\pm$ 0.45 & -1.86 $\pm$ 0.27 & 0.06 $\pm$ 0.14 & 0.15 & 13 & Empirical & \citet{matsunaga2006}\\
$H$& -1.13 $\pm$ 0.36 & -2.35 $\pm$ 0.22 & 0.0 $\pm$ 0.11 & 0.12 & 13 & Empirical & \citet{matsunaga2006}\\
$K_S$& -0.84 $\pm$ 0.35 & -2.6 $\pm$ 0.21 & -0.01 $\pm$ 0.11 & 0.12 & 13& Empirical & \citet{matsunaga2006}\\
$W_{JK_S}$ & -0.46 $\pm$ 0.34 & -3.12 $\pm$ 0.2 & -0.06 $\pm$ 0.1 & 0.11 & 13& Empirical & \citet{matsunaga2006}\\
\hline
\multicolumn{8}{c}{BL Her + W Vir}\\
\hline
$I$ & -0.26 $\pm$ 0.19 & -2.10 $\pm$ 0.06 & 0.04 $\pm$ 0.01 & $-$ & $-$ & Theoretical & \citet{dicriscienzo2007}$^*$\\
$I$ & -0.101 $\pm$ 0.011 & -1.955$\pm$ 0.018 & 0.086 $\pm$ 0.006 & 0.174 & 886 & Theoretical & \citet{marconi2025}\\
$J$ & -1.16 $\pm$ 0.14 & -2.22 $\pm$ 0.07 & -0.19 $\pm$ 0.08 & 0.14 & 33 & Empirical & \citet{matsunaga2006}\\
$J$ & -0.64 $\pm$ 0.13 & -2.29$\pm$ 0.04 & 0.04 $\pm$ 0.01 & $-$ & $-$ & Theoretical & \citet{dicriscienzo2007}\\
$J$ & $-$ & $-$ & -0.318$\pm$0.074 & $-$ & 7 & Empirical & \citet{wielgorski2022}\\
$J$ & -0.498 $\pm$ 0.007 & -2.207$\pm$ 0.012 & 0.076 $\pm$ 0.004 & 0.119 & 886 & Theoretical & \citet{marconi2025}\\
$H$ & -1.51 $\pm$ 0.14 & -2.3 $\pm$ 0.07 & -0.23 $\pm$ 0.08 & 0.14 & 33 & Empirical & \citet{matsunaga2006}\\
$H$ & -0.95 $\pm$ 0.06 & -2.34$\pm$ 0.02 & 0.06 $\pm$ 0.01 & $-$ & $-$ & Theoretical & \citet{dicriscienzo2007}\\
$H$ & $-$ & $-$ & -0.196$\pm$0.070 & $-$ & 7 & Empirical & \citet{wielgorski2022}\\
$H$ & -0.800 $\pm$ 0.004 & -2.426$\pm$ 0.008 & 0.071 $\pm$ 0.003 & 0.073 & 886 & Theoretical & \citet{marconi2025}\\
$K_S$ & -1.39 $\pm$ 0.14 & -2.34 $\pm$ 0.07 & -0.15 $\pm$ 0.08 & 0.14 & 33& Empirical & \citet{matsunaga2006}\\
$K$ & -0.97 $\pm$ 0.06 & -2.38$\pm$ 0.02 & 0.06 $\pm$ 0.01 & $-$ & $-$ & Theoretical & \citet{dicriscienzo2007}\\
$K_S$ & $-$ & $-$ & -0.228$\pm$0.076 & $-$ & 7 & Empirical & \citet{wielgorski2022}\\
$K$ & -0.844 $\pm$ 0.004 & -2.455$\pm$ 0.007 & 0.071 $\pm$ 0.002 & 0.069 & 886 & Theoretical & \citet{marconi2025}\\

$G$ & 0.385 $\pm$ 0.015 & -1.716$\pm$ 0.023 & 0.111 $\pm$ 0.008 & 0.233 & 886 & Theoretical & \citet{marconi2025}\\
$G_{BP}$ & 0.672 $\pm$ 0.019 & -1.524$\pm$ 0.029 & 0.141 $\pm$ 0.010 & 0.281 & 886 & Theoretical & \citet{marconi2025}\\
$G_{RP}$ & -0.056 $\pm$ 0.011 & -1.92$\pm$ 0.019 & 0.086 $\pm$ 0.006 & 0.183 & 886 & Theoretical & \citet{marconi2025}\\

$W_{VI}$ & -1.16 $\pm$ 0.07 & -2.43$\pm$ 0.02 & 0.04 $\pm$ 0.01 & $-$ & $-$ & Theoretical & \citet{dicriscienzo2007}\\
$W_{VI}$ & -1.125 $\pm$ 0.003 & -2.526$\pm$ 0.005 & 0.040$\pm$ 0.002 & 0.047 & 886 & Theoretical & \citet{marconi2025}\\


$W_{JK_S}$ & -1.55 $\pm$ 0.16 & -2.43 $\pm$ 0.07 & -0.12 $\pm$ 0.09 & 0.15 & 33& Empirical & \citet{matsunaga2006}\\
$W_{JK}$ & -1.15 $\pm$ 0.06 & -2.60$\pm$ 0.02 & 0.06 $\pm$ 0.01 & $-$ & $-$ & Theoretical & \citet{dicriscienzo2007}\\
$W_{JK_S}$ & $-$ & $-$ & -0.181$\pm$0.076 & $-$ & 7 & Empirical & \citet{wielgorski2022}\\
$W_{JK}$ & -1.084 $\pm$ 0.003 & -2.627$\pm$ 0.006& 0.067 $\pm$ 0.002 & 0.044 & 886 & Theoretical & \citet{marconi2025}\\

$W_{G_{BP}, G_{RP}}^G$ & -0.992 $\pm$ 0.003 & -2.466$\pm$ 0.006 & 0.006 $\pm$ 0.002 & 0.059 & 886 & Theoretical & \citet{marconi2025}\\
$W_{gi \rm{LSST}}$ & -0.250 $\pm$ 0.003 & -2.561 $\pm$ 0.005 & -0.013 $\pm$ 0.002 & 0.046 & 886 & Theoretical & \citet{marconi2025}\\
$W_{iz\rm{LSST}}$ & 0.296 $\pm$ 0.006 & -2.315 $\pm$ 0.01 & 0.089 $\pm$ 0.004 & 0.099 & 886 & Theoretical & \citet{marconi2025}\\
$W_{gy\rm{LSST}}$ & 0.048 $\pm$ 0.004 & -2.417 $\pm$ 0.007 & 0.039$\pm$ 0.003 & 0.072 & 886 & Theoretical & \citet{marconi2025}\\

\hline
\multicolumn{8}{c}{T2CEP}\\
\hline
$B$ & 1.35 $\pm$ 0.00 & -1.69 $\pm$ 0.05 & 0.35 $\pm$ 0.00 & $-$ & 40 & Empirical & \citet{nemec1994}\\
$B$ & 0.68 $\pm$ 0.25 & -1.67 $\pm$ 0.14 & 0.19 $\pm$ 0.14 & 0.41 & 42 & Empirical & \citet{ngeow2022}\\

$V$ & 0.84 $\pm$ 0.00 & -1.93 $\pm$ 0.05 & 0.32 $\pm$ 0.00 & $-$ & 32 & Empirical & \citet{nemec1994}\\
$V$ & 0.28 $\pm$ 0.20 & -1.90 $\pm$ 0.10 & 0.10 $\pm$ 0.11 & 0.31 & 37 & Empirical & \citet{ngeow2022}\\

$I$ & -0.41 $\pm$ 0.15 & -2.09 $\pm$ 0.09 & -0.01 $\pm$ 0.08 & 0.24 & 41 & Empirical & \citet{ngeow2022}\\

$g$ & -0.22 $\pm$ 0.22 & -1.61 $\pm$ 0.11 & -0.09 $\pm$ 0.12 & 0.38 & 55 & Empirical & \citet{ngeow2022}\\
$r$ & -0.29 $\pm$ 0.17 & -1.83 $\pm$ 0.08 & -0.03 $\pm$ 0.10 & 0.30 & 55 & Empirical & \citet{ngeow2022}\\
$i$ & -0.38 $\pm$ 0.17 & -1.94 $\pm$ 0.08 & -0.07 $\pm$ 0.10 & 0.28 & 41 & Empirical & \citet{ngeow2022}\\

$J$ & -1.06 $\pm$ 0.13 & -2.22 $\pm$ 0.05 & -0.12 $\pm$ 0.07 & 0.15 & 46 & Empirical & \citet{matsunaga2006}\\
$J$ & -0.83 $\pm$ 0.10 & -2.23 $\pm$ 0.04 & -0.00 $\pm$ 0.05 & 0.13 & 45 & Empirical & \citet{ngeow2022}\\
$H$& -1.37 $\pm$ 0.12 & -2.33 $\pm$ 0.05 & -0.15 $\pm$ 0.07 & 0.14 & 46 & Empirical & \citet{matsunaga2006}\\
$H$ & -1.14 $\pm$ 0.08 & -2.35 $\pm$ 0.03 & -0.05 $\pm$ 0.04 & 0.10 & 43 & Empirical & \citet{ngeow2022}\\
$K_S$ & -1.27 $\pm$ 0.11 & -2.4 $\pm$ 0.05 & -0.1 $\pm$ 0.06 & 0.14 & 46& Empirical & \citet{matsunaga2006}\\
$K$ & -1.10 $\pm$ 0.08 & -2.41 $\pm$ 0.03 & -0.00 $\pm$ 0.04 & 0.10 & 48 & Empirical & \citet{ngeow2022}\\
$W_{JK_S}$& -1.42 $\pm$ 0.13 & -2.52 $\pm$ 0.05 & -0.08 $\pm$ 0.07 & 0.16 & 46& Empirical & \citet{matsunaga2006}\\
\hline
\end{tabular}}
\label{tab:PLZ}
\end{table*}

\begin{figure*}
\centering
\includegraphics[scale = 1]{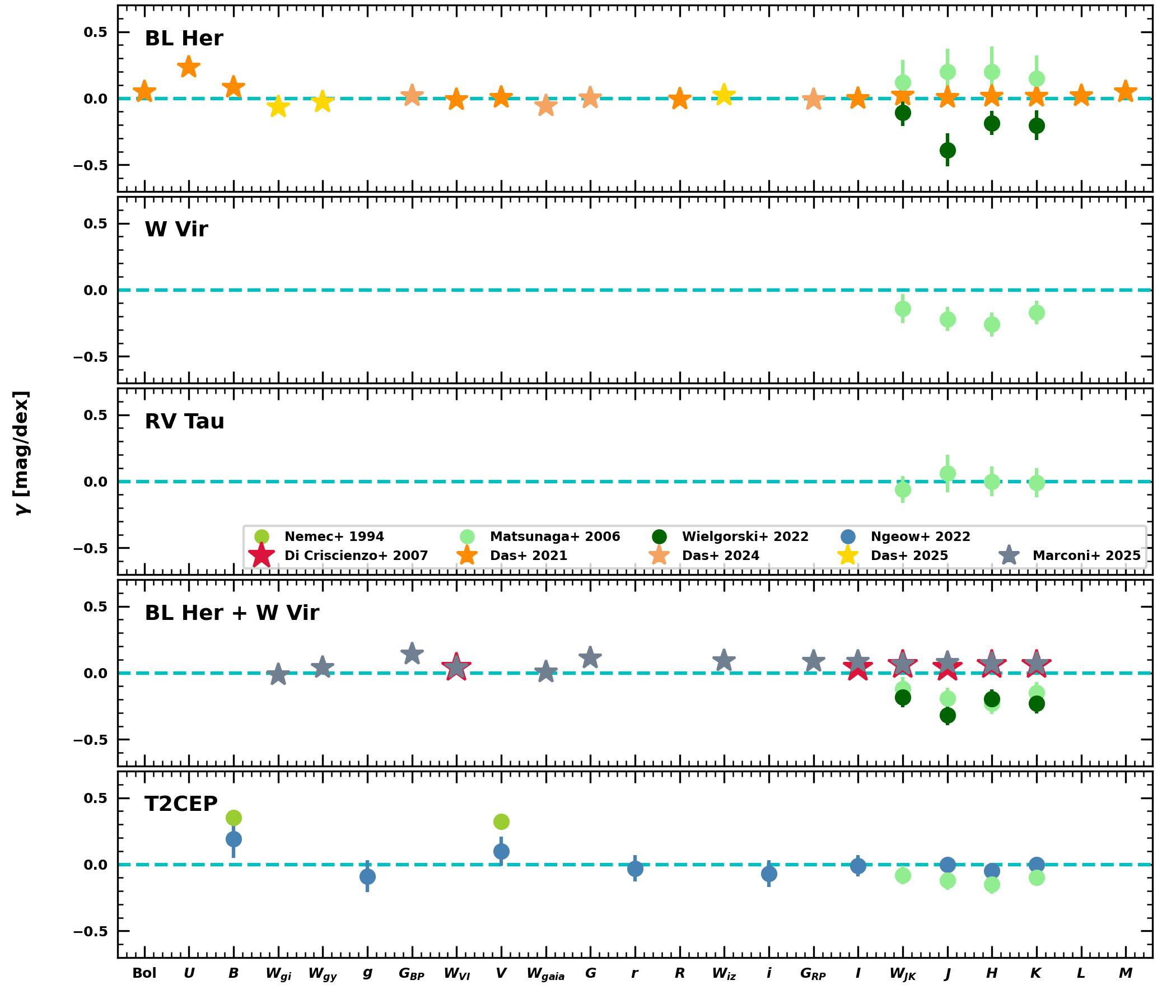}
\caption{Empirical (in circles) and theoretical (in star-shaped symbols) estimates of the T2CEP metallicity dependence $\gamma$ from the literature across different passbands. The top three panels present results from the individual T2CEP subclasses- BL~Her, W~Vir and RV~Tau stars. The bottom two panels demonstrate the effect of metallicity when different subclasses are combined- BL~Her+W~Vir and T2CEP (BL~Her+W~Vir+RV~Tau), respectively. Passbands include the bolometric (Bol), Johnson-Cousins-Glass ($UBVRIJHKLMW_{VI}W_{JK}$), ZTF ($gri$), Gaia ($GG_{BP}G_{RP}W_{gaia}$) and the Rubin--LSST ($W_{gi}, W_{gy}, W_{iz}$) filters. The x-axis is in increasing order of the central effective wavelengths ($\lambda_{\rm eff}$) of the respective passbands as provided by the SVO Filter Profile Service \citep{rodrigo2012, rodrigo2020}.}
\label{fig:gamma}
\end{figure*}

Almost a decade later, the study of the metallicity dependence (or lack thereof) was re-ignited in a series of papers by \citet{matsunaga2006, matsunaga2009, matsunaga2011}. The photometric data of T2CEPs analysed in \citet{matsunaga2006} were obtained from the Infrared Survey Facility (IRSF) 1.4-m telescope and the Simultaneous 3-Colour
Imager for Unbiased Survey (SIRIUS) near-infrared camera \citep{nagashima1999, nagayama2003} while the parameters for the globular clusters $-$ the metallicity [Fe/H], the colour excess E(B-V) and the magnitude of the horizontal branch V(HB) were
adopted from \citet{harris1996}. For the reddening corrections, $R_V=3.1$ was used along with extinction law from \citet{cardelli1989}. The distance moduli adopted were based on the magnitudes of horizontal branches of the clusters \citep[using the relation from][]{gratton2003}, which were then used to obtain the absolute magnitudes of the T2CEPs. While \citet{matsunaga2006} found the metallicity contribution ($\gamma = -0.10\pm0.06$) towards the $K$-band $PL$ relation from 46 T2CEPs in 26 Galactic Globular Clusters to be hardly significant, \citet{matsunaga2011} presented evidence of different $PL$ relations obeyed by T2CEPs when three different systems are considered (globular clusters, LMC and SMC). Using the method of estimating the difference in the distance moduli between LMC and SMC, they also concluded that the absolute magnitudes of W~Vir stars are free of metallicity effects while BL~Her stars are not. In a more recent empirical study, \citet{wielgorski2022} used photometric data of T2CEPs from the 0.81 m InfraRed Imaging Survey (IRIS) telescope \citep{hodapp2010, ramolla2016} which were corrected for interstellar extinction using reddening maps from \citet{schlafly2011} to obtain the colour excess E(B-V) for each star, with $R_V=3.1$ and the reddening law from \citet{cardelli1989} and \citet{odonnell11994}. The distances to the individual T2CEPs were obtained from Gaia EDR3 parallaxes \citep{lindegren2021} while the spectroscopic metallicity determinations [Fe/H] for a fraction of Milky Way T2CEPs were taken from \citet{maas2007}. They obtained a metallicity effect of $-0.2$ mag dex$^{-1}$, albeit with the caveat that the sample size is rather small with a total of 7 T2CEPs only, BL~Her and W~Vir stars included. In yet another recent empirical study, \citet{ngeow2022} provided new $PLZ$ relations for 37 T2CEPs in 18 globular clusters in the $gri$ bands using data from the Zwicky Transient Facility \citep[ZTF,][]{bellm2019} and updated $BVIJHK$ band relations for 58 T2CEPS in 24 globular clusters. The distances to the host globular clusters were obtained from \citet{baumgardt2021}. For the $gri$ bands, reddening map from \citet{green2019} was used to obtain the colour excess towards each star, followed by extinction correction relations also from \citet{green2019}. Individual reddening values for each T2CEP in the $BVIJHK$ bands were derived from the \citet{schlegel1998} dust map, with extinction corrections computed using the relations of \citet{schlafly2011, green2019}. The metallicities [Fe/H] for the host globular clusters were obtained from the GlObular clusTer Homogeneous Abundances Measurements (GOTHAM) survey \citet{dias2015, dias2016a, dias2016b, vasquez2018}. Except for the $B$-band, the metallicity effect towards the $PL$ relations in all the other bands ($gri$ and $VIJHK$) were found to be negligible within reported uncertainties. We note that different studies adopt varying methods for deriving extinction corrections, metallicities and distances. Consequently, some degree of inhomogeneity may arise when comparing the metallicity dependence of PL relations across empirical works. Moreover, assuming that the metallicities of T2CEPs in globular clusters are identical to the mean metallicities of their host clusters can introduce substantial uncertainty in the metallicity term of the $PLZ$ relation.

On the theoretical front, \citet{dicriscienzo2007} computed pulsation models following \citet{bono1994}, with identical nonlinear, nonlocal, time-dependent convective hydrodynamics and the same equation-of-state and opacity prescriptions. In particular, opacity compilations for temperatures higher than 10,000 K were used from \citet{iglesias1996} and for lower temperatures from \citet{alexander1994}. The bolometric light curves were transformed into respective passbands by adopting bolometric corrections and temperature-color transformations from \citet{castelli1997b, castelli1997a}. They thereby provided theoretical relations over multiple wavelengths for the short-period T2CEPs. However, for the purpose of this review, we follow the same convention for BL~Her stars as those with pulsation periods between 1 and 4 days while for W~Vir stars as those with periods between 4 and 20 days for a uniform comparison throughout the literature. Therefore, although \citet{dicriscienzo2007} presents the results for BL~Her models, we include the study in the BL~Her+W~Vir category for this review since the models span the range of pulsation periods up to 8 days. In a series of papers, \citet{das2021,das2024,das2025b} presented theoretical period relations for BL~Her stars across multiple wavelength bands, including Johnson–Cousins–Glass ($UBVRIJHKLL'M$), Gaia ($GG_{BP} G_{RP}$), and for the first time, in the Rubin-LSST\footnote{\url{https://rubinobservatory.org/}} ($ugrizy$) passbands. The relations were derived from a fine grid of non-linear convective models computed with the Radial Stellar Pulsation \citep[\textsc{rsp},][]{smolec2008,paxton2019} tool within the \textit{Modules for Experiments in Stellar Astrophysics} \citep[\textsc{mesa},][]{paxton2011,paxton2013,paxton2015,paxton2018,paxton2019,jermyn2023} software suite. \textsc{mesa-rsp} follows the turbulent convection formulation of \citet{kuhfuss1986} and the stellar pulsation treatment of \citet{smolec2008}. OPAL opacities \citep{iglesias1996}, supplemented at low temperatures by \citet{ferguson2005}, were adopted. Bolometric light curves were converted to the Johnson–Cousins–Glass ($UBVRIJHKLL'M$) system using the pre-computed bolometric correction table of \citet{lejeune1998}, while transformations to the Gaia ($G,G_{BP},G_{RP}$) and Rubin–LSST $ugrizy$ passbands were performed using those from MIST\footnote{\url{https://waps.cfa.harvard.edu/MIST/index.html}} (\textsc{mesa} Isochrones \& Stellar Tracks) packaged model grids. Theoretical $PL$, period-Wesenheit ($PW$) and period-radius ($PR$) relations showed good agreement with most empirical relations, with no significant metallicity dependence in the $PL$ (except for $U$ and $B$ bands) and $PR$ relations. Finally, in the most recent theoretical study of T2CEPs, \citet{marconi2025} presented an extensive set of models for BL~Her and short period W~Vir (up to 10 days) stars, following the methodology as in \citet{dicriscienzo2007} but with opacity tables as was used in \citet{desomma2024}. In particular, the radiative Rosseland opacity was taken from \citet{iglesias1996} for temperatures higher than $\log(T ) = 4.0$ and from \citet{ferguson2005} for lower temperatures. They predicted light and radial velocity curves and thereby obtained $PLZ$ and $PWZ$ relations across a wide range of wavelengths, including $IJHK$, Gaia, VISTA and Rubin-LSST.

Table~\ref{tab:PLZ} presents, to the best of our knowledge, a comprehensive compilation of all available $PLZ$ relations across multiple wavelengths reported in the literature, which are also illustrated in Fig.~\ref{fig:gamma}. While the set of studies included here can be regarded as comprehensive, the coverage of photometric passbands may nevertheless remain incomplete and the interested reader is encouraged to refer to the original works for more details. This is especially true for the case of theoretical studies where we have included results for the most commonly used passbands and/or for the simplest convection parameters only. In particular, the $PLZ$ relations from \citet{dicriscienzo2007} included here were obtained for the particular case of mixing length parameter 1.5. For the theoretical relations from \citet{das2021,das2024,das2025b}, we have only included the results computed using the simplest convection parameter set while the papers also contain results as a function of different convection parameter sets. In addition, we only highlight the relations involving the Wesenheit indices $W(i,g-i)$, $W(z,i-z)$, and $W(y,g-y)$ in the Rubin-LSST passbands, owing to their minimal sensitivity to metallicity. These filter combinations are therefore considered optimal for future Rubin–LSST observations of T2CEPs, supporting their use as reliable standard candles. The interested reader is referred to \citet{das2025b} for individual $PLZ$ relations in the Rubin-LSST $ugrizy$ passbands. Lastly, \citet{marconi2025} also provides $PLZ$ relations for the VISTA $JYK_s$ bands, in addition to the passbands included here in this review. Note that while \citet{dicriscienzo2007} and \citet{marconi2025} used the convection formulation outlined in \citet{stellingwerf1982a, stellingwerf1982b}, \citet{das2021,das2024,das2025b} used \textsc{mesa-rsp,} which follows the turbulent convection theory from \citet{kuhfuss1986}. Therefore, the negligible effect of metallicity towards the $PL$ relations of T2CEPs appears promising and remains robust across various theoretical stellar pulsation codes, even when different theories of convection are employed.

\section{Applications in Distance Scale and Precision Cosmology}

Classical Cepheids have been extensively used in the first rung of the distance ladder scale of the SH0ES program \citep[Supernovae and $H_0$ for the Equation of State of dark energy,][]{riess2009} for the subsequent estimation of the local Hubble constant. As an alternative route, the Carnegie‐Chicago Hubble Program \citep[CCHP,][]{beaton2016} uses the independent approach of using only Population~II distance indicators (RR~Lyraes and TRGBs) to estimate the same. The geometric calibration of the $PL$ relations of classical pulsators therefore not only serve in estimating extragalactic distances but is also crucial to provide direct measurements of the $H_0$ using the distance versus redshift approach as opposed to the indirect measurement from an early Universe calibration of the $\Lambda$ cold dark matter model ($\Lambda$CDM) model-dependent inference from the Planck CMB observations \citep{planck2020}.  The present values of the Hubble constant derived from the early Universe approach is $H_0 = 67.4 \pm 0.5$ km $s^{-1}$ Mpc \citep{planck2020} while from the local Universe approach is $H_0 = 73.18 \pm 0.88$ km $s^{-1}$ Mpc \citep{riess2022, bhardwaj2023, riess2024, riess2025}, resulting in the Hubble tension with a $\sim 6\sigma$ discrepancy \citep[see also][]{verde2019}. With the recent improvements in the precision of $H_0$ measurements and its critical role in testing extensions to $\Lambda$CDM, addressing and reducing previously unrecognized sources of uncertainty and bias in $H_0$ has become increasingly crucial.

In this context, it is essential to validate distances derived from classical Cepheids by employing independent populations of pulsating variables. T2CEPs are particularly promising for this purpose due to their minimal metallicity dependence in the $PL$ relations across different passbands, as discussed in the previous section. Several theoretical and empirical studies have also shown that RR~Lyraes and T2CEPs follow a common $PL$ relation, especially at near-infrared bands \citep[for example,][]{majaess2010, bhardwaj2017c, braga2020, das2025b, lengen2025}. Consequently, when combined with RR~Lyrae stars and the TRGB, they could serve as an alternative to classical Cepheids for extragalactic distance measurements. This methodology not only allows for rigorous cross-validation of classical Cepheid-based distances but also extends the applicability of pulsating variables as robust standard candles across a range of stellar populations and metallicity regimes, ultimately contributing to more accurate $H_0$ determinations \citep[see e.g.][]{beaton2016}.

Despite their promise, T2CEPs currently face a few challenges when applied to precision cosmology. The intrinsically lower luminosity of T2CEPs restricts their effective use as distance indicators to approximately only a few Mpc with present-day instrumentation and the population of known T2CEPs beyond the Local Group therefore remains small. Studies of T2CEPs in the Milky Way have largely focused on globular clusters and the Galactic bulge \citep[e.g.,][]{bhardwaj2017b, bhardwaj2022, narloch2025,cruzreyes2025}. Investigations in the Magellanic Clouds have examined their luminosities and $PL$ relations in detail \citep[e.g.,][]{bhardwaj2017a, groenewegen2017,sicignano2024}. Applications to nearby Local Group galaxies beyond the Magellanic Clouds remain limited, with only a few efforts addressing their potential as extragalactic distance indicators \citep[e.g.,][]{majaess2009}. The metallicity calibration of T2CEPs therefore remains less well-established compared to that of Classical Cepheids. However, with the advent of next-generation surveys such as Rubin-LSST and the Roman Space Telescope\footnote{\url{https://roman.gsfc.nasa.gov/}}, the potential applications of T2CEPs are expected to increase substantially. Although not suitable for direct determinations of the Hubble constant $H_0$, T2CEPs could provide important cross-checks on Cepheid-based distances, especially in galaxies hosting both Cepheid and T2CEP populations and in anchoring TRGB distances, which are used in recent independent measurements of $H_0$. Their relative insensitivity to metallicity even at optical wavelengths, combined with their prevalence in older stellar populations, complements the younger, more metallicity-dependent classical Cepheid distance scale.

\section{Summary}

This work provides a comprehensive review of the effect of metallicity on the $PL$ relation of T2CEPs. Most empirical \citep{nemec1994, matsunaga2006, ngeow2022} and theoretical \citep{dicriscienzo2007, das2021, das2024, das2025b, marconi2025} studies to date suggest that metallicity has a weak or negligible effect on the T2CEP $PL$ relation across a broad range of wavelengths, with the exception of the shorter-wavelength $U$ and $B$ bands \citep{nemec1994, das2021}. For distance scale applications, this potentially provides a significant advantage over other commonly used classical pulsators, such as RR~Lyraes and classical Cepheids, which are known to exhibit strong metallicity effects, particularly at shorter wavelengths. In addition, RR~Lyraes and T2CEPs, which have been shown to follow similar $PL$ relations \citep{majaess2010, bhardwaj2017c, braga2020, das2025b, lengen2025}, can be combined with the TRGB to provide an independent cross-check of Cepheid-based distances and serve as a complementary Population~II distance scale.

Despite significant progress, the field of T2CEPs still presents many open challenges. On the theoretical side, current stellar pulsation codes are able to reliably reproduce the behavior of BL~Her and short-period W~Vir stars, but they face substantial difficulties in modeling long-period W Vir and RV Tau variables. These complications arise from the unstable outer layers of such stars, where the radiation-diffusion approximation breaks down and pulsation-driven mass loss may also need to be included to adequately describe their variability \citep{smolec2016, paxton2019}. Empirical $PLZ$ relations for T2CEPs currently show good agreement with theoretical predictions, even when the latter are computed with different convection parameter sets \citep{das2021, das2024}. This consistency may partly reflect the limited precision of current observations, which remain insufficient to decisively distinguish among models based on different convection treatments. The advent of forthcoming large-scale surveys is expected to deliver much higher precision data, making it essential to calibrate convection parameters for T2CEP models, as has recently been done for RR~Lyrae stars \citep{kovacs2023, kovacs2024}. Beyond calibrating $PLZ$ relations, it is equally important to test whether theoretical light curves can successfully reproduce observed pulsation periods and detailed light-curve morphologies. Such comparisons are vital for ensuring that models capture the underlying physics throughout the entire pulsation cycle, rather than only at mean-light, as recently demonstrated by \citet{das2025a}. Finally, the development of a self-consistent framework that simultaneously incorporates stellar evolution and pulsation physics represents an important step forward, as illustrated by the work of \citet{marconi2025}. On the empirical side, current instrumentation already provides high-quality photometric data for T2CEPs. Nevertheless, a homogeneous determination of metallic abundances from high-resolution spectroscopy across a large sample remains indispensable for achieving a more precise characterization of metallicity effects. Thanks to the Gaia space mission which has provided geometric distance measurements for several dozen field T2CEPs, it will soon be possible to calibrate the $PLZ$ relations of these stars with unprecedented precision and accuracy, once homogeneous photometry and metallicity data become available.

In summary, T2CEPs are emerging as reliable and complementary distance indicators, owing to the minimal impact of metallicity on their $PL$ relations. Their presence in older stellar populations and the shared $PL$ relation with RR~Lyraes provide an independent distance scale that can cross-validate classical Cepheid distances. Forthcoming large-scale photometric and spectroscopic surveys will enable more precise calibrations of their $PLZ$ relations, detailed comparisons of theoretical and observed light curves, and improved assessments of metallicity effects. As a result, T2CEPs are poised to play an increasingly important role in refining the extragalactic distance scale and constraining $H_0$.

\section*{Conflict of Interest Statement}
The authors declare that the research was conducted in the absence of any commercial or financial relationships that could be construed as a potential conflict of interest. The author(s) declare that they were an editorial board member of Frontiers, at the time of submission. This had no impact on the peer review process and the final decision.

\section*{Author Contributions}

SD: Writing – original draft, Writing – review \& editing. AB: Writing – review \& editing. MM: Writing – review \& editing.

\section*{Funding}
This research was supported by the International Space Science Institute (ISSI) in Bern/Beijing through ISSI/ISSI-BJ International Team project ID \#24-603 - “EXPANDING Universe” (EXploiting Precision AstroNomical Distance INdicators in the Gaia Universe). SD acknowledges the KKP-137523 `SeismoLab' \'Elvonal grant of the Hungarian Research, Development and Innovation Office (NKFIH). A.B. thanks the funding from the Anusandhan National Research Foundation (ANRF) under the Prime Minister Early Career Research Grant scheme (ANRF/ECRG/2024/000675/PMS).

\section*{Acknowledgments}
The authors are grateful to the referee for useful suggestions that improved the quality of the manuscript. This research has made use of NASA’s Astrophysics Data System.


\begin{thebibliography}{118}
\providecommand{\natexlab}[1]{#1}
\expandafter\ifx\csname urlstyle\endcsname\relax
  \providecommand{\doi}[1]{doi:\discretionary{}{}{}#1}\else
  \providecommand{\doi}{doi:\discretionary{}{}{}\begingroup
  \urlstyle{rm}\Url}\fi
\providecommand{\selectlanguage}[1]{\relax}
\providecommand{\bibAnnoteFile}[1]{%
  \IfFileExists{#1}{\begin{quotation}\noindent\textsc{Key:} #1\\
  \textsc{Annotation:}\ \input{#1}\end{quotation}}{}}
\providecommand{\bibAnnote}[2]{%
  \begin{quotation}\noindent\textsc{Key:} #1\\
  \textsc{Annotation:}\ #2\end{quotation}}

\bibitem[{{Alcock} et~al.(1998){Alcock}, {Allsman}, {Alves}, {Axelrod},
  {Becker}, {Bennett} et~al.}]{alcock1998}
{Alcock}, C., {Allsman}, R.~A., {Alves}, D.~R., {Axelrod}, T.~S., {Becker}, A.,
  {Bennett}, D.~P., et~al. (1998).
\newblock {The MACHO Project LMC Variable Star Inventory. VII. The Discovery of
  RV Tauri Stars and New Type II Cepheids in the Large Magellanic Cloud}.
\newblock \emph{AJ} 115, 1921--1933.
\newblock \doi{10.1086/300317}
\input{alcock1998}

\bibitem[{{Alexander} and {Ferguson}(1994)}]{alexander1994}
{Alexander}, D.~R. and {Ferguson}, J.~W. (1994).
\newblock {Low-Temperature Rosseland Opacities}.
\newblock \emph{ApJ} 437, 879.
\newblock \doi{10.1086/175039}
\input{alexander1994}

\bibitem[{{Arp}(1955)}]{arp1955}
{Arp}, H.~C. (1955).
\newblock {Cepheids of period greater than 1 day in globular clusters.}
\newblock \emph{AJ} 60, 1--17.
\newblock \doi{10.1086/107091}
\input{arp1955}

\bibitem[{{Baade}(1944)}]{baade1944}
{Baade}, W. (1944).
\newblock {The Resolution of Messier 32, NGC 205, and the Central Region of the
  Andromeda Nebula.}
\newblock \emph{ApJ} 100, 137.
\newblock \doi{10.1086/144650}
\input{baade1944}

\bibitem[{{Baade}(1956)}]{baade1956}
{Baade}, W. (1956).
\newblock {The Period-Luminosity Relation of the Cepheids}.
\newblock \emph{PASP} 68, 5.
\newblock \doi{10.1086/126870}
\input{baade1956}

\bibitem[{{Balog} et~al.(1997){Balog}, {Vinko}, and {Kaszas}}]{balog1997}
{Balog}, Z., {Vinko}, J., and {Kaszas}, G. (1997).
\newblock {Baade-Wesselink Radius Determination of Type II Cepheids}.
\newblock \emph{AJ} 113, 1833.
\newblock \doi{10.1086/118394}
\input{balog1997}

\bibitem[{{Baumgardt} and {Vasiliev}(2021)}]{baumgardt2021}
{Baumgardt}, H. and {Vasiliev}, E. (2021).
\newblock {Accurate distances to Galactic globular clusters through a
  combination of Gaia EDR3, HST, and literature data}.
\newblock \emph{MNRAS} 505, 5957--5977.
\newblock \doi{10.1093/mnras/stab1474}
\input{baumgardt2021}

\bibitem[{{Beaton} et~al.(2016){Beaton}, {Freedman}, {Madore}, {Bono},
  {Carlson}, {Clementini} et~al.}]{beaton2016}
{Beaton}, R.~L., {Freedman}, W.~L., {Madore}, B.~F., {Bono}, G., {Carlson},
  E.~K., {Clementini}, G., et~al. (2016).
\newblock {The Carnegie-Chicago Hubble Program. I. An Independent Approach to
  the Extragalactic Distance Scale Using Only Population II Distance
  Indicators}.
\newblock \emph{ApJ} 832, 210.
\newblock \doi{10.3847/0004-637X/832/2/210}
\input{beaton2016}

\bibitem[{{Bellm} et~al.(2019){Bellm}, {Kulkarni}, {Graham}, {Dekany}, {Smith},
  {Riddle} et~al.}]{bellm2019}
{Bellm}, E.~C., {Kulkarni}, S.~R., {Graham}, M.~J., {Dekany}, R., {Smith},
  R.~M., {Riddle}, R., et~al. (2019).
\newblock {The Zwicky Transient Facility: System Overview, Performance, and
  First Results}.
\newblock \emph{PASP} 131, 018002.
\newblock \doi{10.1088/1538-3873/aaecbe}
\input{bellm2019}

\bibitem[{{Bersier} et~al.(1997){Bersier}, {Burki}, and {Kurucz}}]{bersier1997}
{Bersier}, D., {Burki}, G., and {Kurucz}, R.~L. (1997).
\newblock {Fundamental parameters of Cepheids. IV. Radii and luminosities.}
\newblock \emph{A\&A} 320, 228--236
\input{bersier1997}

\bibitem[{{Bhardwaj}(2020)}]{bhardwaj2020}
{Bhardwaj}, A. (2020).
\newblock {High-precision distance measurements with classical pulsating
  stars}.
\newblock \emph{Journal of Astrophysics and Astronomy} 41, 23.
\newblock \doi{10.1007/s12036-020-09640-z}
\input{bhardwaj2020}

\bibitem[{{Bhardwaj} et~al.(2017{\natexlab{a}}){Bhardwaj}, {Kanbur}, {Marconi},
  {Rejkuba}, {Singh}, and {Ngeow}}]{bhardwaj2017c}
{Bhardwaj}, A., {Kanbur}, S.~M., {Marconi}, M., {Rejkuba}, M., {Singh}, H.~P.,
  and {Ngeow}, C.-C. (2017{\natexlab{a}}).
\newblock {A comparative study of multiwavelength theoretical and observed
  light curves of Cepheid variables}.
\newblock \emph{MNRAS} 466, 2805--2824.
\newblock \doi{10.1093/MNRAS/stw3256}
\input{bhardwaj2017c}

\bibitem[{{Bhardwaj} et~al.(2022){Bhardwaj}, {Kanbur}, {Rejkuba}, {Marconi},
  {Catelan}, {Ripepi} et~al.}]{bhardwaj2022}
{Bhardwaj}, A., {Kanbur}, S.~M., {Rejkuba}, M., {Marconi}, M., {Catelan}, M.,
  {Ripepi}, V., et~al. (2022).
\newblock {Near-infrared observations of RR Lyrae and Type II Cepheid variables
  in the metal-rich bulge globular cluster NGC 6441}.
\newblock \emph{A\&A} 668, A59.
\newblock \doi{10.1051/0004-6361/202244728}
\input{bhardwaj2022}

\bibitem[{{Bhardwaj} et~al.(2017{\natexlab{b}}){Bhardwaj}, {Macri}, {Rejkuba},
  {Kanbur}, {Ngeow}, and {Singh}}]{bhardwaj2017a}
{Bhardwaj}, A., {Macri}, L.~M., {Rejkuba}, M., {Kanbur}, S.~M., {Ngeow}, C.-C.,
  and {Singh}, H.~P. (2017{\natexlab{b}}).
\newblock {Large Magellanic Cloud Near-infrared Synoptic Survey. IV. Leavitt
  Laws for Type II Cepheid Variables}.
\newblock \emph{AJ} 153, 154.
\newblock \doi{10.3847/1538-3881/aa5e4f}
\input{bhardwaj2017a}

\bibitem[{{Bhardwaj} et~al.(2017{\natexlab{c}}){Bhardwaj}, {Rejkuba},
  {Minniti}, {Surot}, {Valenti}, {Zoccali} et~al.}]{bhardwaj2017b}
{Bhardwaj}, A., {Rejkuba}, M., {Minniti}, D., {Surot}, F., {Valenti}, E.,
  {Zoccali}, M., et~al. (2017{\natexlab{c}}).
\newblock {Galactic bulge population II Cepheids in the VVV survey:
  period-luminosity relations and a distance to the Galactic centre}.
\newblock \emph{A\&A} 605, A100.
\newblock \doi{10.1051/0004-6361/201730841}
\input{bhardwaj2017b}

\bibitem[{{Bhardwaj} et~al.(2023){Bhardwaj}, {Riess}, {Catanzaro}, {Trentin},
  {Ripepi}, {Rejkuba} et~al.}]{bhardwaj2023}
{Bhardwaj}, A., {Riess}, A.~G., {Catanzaro}, G., {Trentin}, E., {Ripepi}, V.,
  {Rejkuba}, M., et~al. (2023).
\newblock {High-resolution Spectroscopic Metallicities of Milky Way Cepheid
  Standards and Their Impact on the Leavitt Law and the Hubble Constant}.
\newblock \emph{ApJL} 955, L13.
\newblock \doi{10.3847/2041-8213/acf710}
\input{bhardwaj2023}

\bibitem[{{Bhardwaj} et~al.(2024){Bhardwaj}, {Ripepi}, {Testa}, {Molinaro},
  {Marconi}, {De Somma} et~al.}]{bhardwaj2024}
{Bhardwaj}, A., {Ripepi}, V., {Testa}, V., {Molinaro}, R., {Marconi}, M., {De
  Somma}, G., et~al. (2024).
\newblock {Cepheid Metallicity in the Leavitt Law (C-MetaLL) survey. V. New
  multiband (grizJHK$_{s}$) Cepheid light curves and period-luminosity
  relations}.
\newblock \emph{A\&A} 683, A234.
\newblock \doi{10.1051/0004-6361/202348140}
\input{bhardwaj2024}

\bibitem[{{Bono} et~al.(2020){Bono}, {Braga}, {Fiorentino}, {Salaris},
  {Pietrinferni}, {Castellani} et~al.}]{bono2020}
{Bono}, G., {Braga}, V.~F., {Fiorentino}, G., {Salaris}, M., {Pietrinferni},
  A., {Castellani}, M., et~al. (2020).
\newblock {Evolutionary and pulsation properties of Type II Cepheids}.
\newblock \emph{A\&A} 644, A96.
\newblock \doi{10.1051/0004-6361/202038191}
\input{bono2020}

\bibitem[{{Bono} et~al.(2024){Bono}, {Braga}, and {Pietrinferni}}]{bono2024}
{Bono}, G., {Braga}, V.~F., and {Pietrinferni}, A. (2024).
\newblock {Cepheids as distance indicators and stellar tracers}.
\newblock \emph{A\&Ar} 32, 4.
\newblock \doi{10.1007/s00159-024-00153-0}
\input{bono2024}

\bibitem[{{Bono} et~al.(1997){Bono}, {Caputo}, and {Santolamazza}}]{bono1997}
{Bono}, G., {Caputo}, F., and {Santolamazza}, P. (1997).
\newblock {Evolutionary scenario for metal-poor pulsating stars. I. Type II
  Cepheids.}
\newblock \emph{A\&A} 317, 171--177
\input{bono1997}

\bibitem[{{Bono} and {Stellingwerf}(1994)}]{bono1994}
{Bono}, G. and {Stellingwerf}, R.~F. (1994).
\newblock {Pulsation and Stability of RR Lyrae Stars. I. Instability Strip}.
\newblock \emph{ApJs} 93, 233.
\newblock \doi{10.1086/192054}
\input{bono1994}

\bibitem[{{Braga} et~al.(2020){Braga}, {Bono}, {Fiorentino}, {Stetson},
  {Dall'Ora}, {Salaris} et~al.}]{braga2020}
{Braga}, V.~F., {Bono}, G., {Fiorentino}, G., {Stetson}, P.~B., {Dall'Ora}, M.,
  {Salaris}, M., et~al. (2020).
\newblock {Separation between RR Lyrae and type II Cepheids and their
  importance for a distance determination: the case of omega Cen}.
\newblock \emph{A\&A} 644, A95.
\newblock \doi{10.1051/0004-6361/202039145}
\input{braga2020}

\bibitem[{{Breuval} et~al.(2025){Breuval}, {Anand}, {Anderson}, {Beaton},
  {Bhardwaj}, {Casertano} et~al.}]{breuval2025}
{Breuval}, L., {Anand}, G.~S., {Anderson}, R.~I., {Beaton}, R., {Bhardwaj}, A.,
  {Casertano}, S., et~al. (2025).
\newblock {Converging on the Cepheid Metallicity Dependence: Implications of
  Non-Standard Gaia Parallax Recalibration on Distance Measures}.
\newblock \emph{arXiv e-prints} ,
  arXiv:2507.15936\doi{10.48550/arXiv.2507.15936}
\input{breuval2025}

\bibitem[{{Caputo}(1998)}]{caputo1998}
{Caputo}, F. (1998).
\newblock {Evolution of Population II stars}.
\newblock \emph{A\&Ar} 9, 33--61.
\newblock \doi{10.1007/s001590050014}
\input{caputo1998}

\bibitem[{{Cardelli} et~al.(1989){Cardelli}, {Clayton}, and
  {Mathis}}]{cardelli1989}
{Cardelli}, J.~A., {Clayton}, G.~C., and {Mathis}, J.~S. (1989).
\newblock {The Relationship between Infrared, Optical, and Ultraviolet
  Extinction}.
\newblock \emph{ApJ} 345, 245.
\newblock \doi{10.1086/167900}
\input{cardelli1989}

\bibitem[{{Carson} and {Stothers}(1982)}]{carson1982}
{Carson}, R. and {Stothers}, R. (1982).
\newblock {BL HER stars : theoretical models for field variables.}
\newblock \emph{ApJ} 259, 740--748.
\newblock \doi{10.1086/160210}
\input{carson1982}

\bibitem[{{Castelli} et~al.(1997{\natexlab{a}}){Castelli}, {Gratton}, and
  {Kurucz}}]{castelli1997b}
{Castelli}, F., {Gratton}, R.~G., and {Kurucz}, R.~L. (1997{\natexlab{a}}).
\newblock {(Erratum) Notes on the convection in the ATLAS9 model atmospheres.}
\newblock \emph{A\&A} 324, 432--432
\input{castelli1997b}

\bibitem[{{Castelli} et~al.(1997{\natexlab{b}}){Castelli}, {Gratton}, and
  {Kurucz}}]{castelli1997a}
{Castelli}, F., {Gratton}, R.~G., and {Kurucz}, R.~L. (1997{\natexlab{b}}).
\newblock {Notes on the convection in the ATLAS9 model atmospheres.}
\newblock \emph{A\&A} 318, 841--869
\input{castelli1997a}

\bibitem[{{Catelan} and {Smith}(2015)}]{catelan2015}
{Catelan}, M. and {Smith}, H.~A. (2015).
\newblock \emph{{Pulsating Stars}}
\input{catelan2015}

\bibitem[{{Cruz Reyes} et~al.(2025){Cruz Reyes}, {Anderson}, and
  {Das}}]{cruzreyes2025}
{Cruz Reyes}, M., {Anderson}, R.~I., and {Das}, S. (2025).
\newblock {Variable stars in Galactic globular clusters: II. Population II
  Cepheids}.
\newblock \emph{A\&A} 695, A164.
\newblock \doi{10.1051/0004-6361/202453137}
\input{cruzreyes2025}

\bibitem[{{Das} et~al.(2021){Das}, {Kanbur}, {Smolec}, {Bhardwaj}, {Singh}, and
  {Rejkuba}}]{das2021}
{Das}, S., {Kanbur}, S.~M., {Smolec}, R., {Bhardwaj}, A., {Singh}, H.~P., and
  {Rejkuba}, M. (2021).
\newblock {A theoretical framework for BL Her stars - I. Effect of metallicity
  and convection parameters on period-luminosity and period-radius relations}.
\newblock \emph{MNRAS} 501, 875--891.
\newblock \doi{10.1093/mnras/staa3694}
\input{das2021}

\bibitem[{{Das} et~al.(2024){Das}, {Moln{\'a}r}, {Kanbur}, {Joyce}, {Bhardwaj},
  {Singh} et~al.}]{das2024}
{Das}, S., {Moln{\'a}r}, L., {Kanbur}, S.~M., {Joyce}, M., {Bhardwaj}, A.,
  {Singh}, H.~P., et~al. (2024).
\newblock {A theoretical framework for BL Her stars. II. New period-luminosity
  relations in Gaia passbands}.
\newblock \emph{A\&A} 684, A170.
\newblock \doi{10.1051/0004-6361/202348280}
\input{das2024}

\bibitem[{{Das} et~al.(2025{\natexlab{a}}){Das}, {Moln{\'a}r}, {Kov{\'a}cs},
  {Smolec}, {Joyce}, {Kanbur} et~al.}]{das2025a}
{Das}, S., {Moln{\'a}r}, L., {Kov{\'a}cs}, G.~B., {Smolec}, R., {Joyce}, M.,
  {Kanbur}, S.~M., et~al. (2025{\natexlab{a}}).
\newblock {A theoretical framework for BL Her stars: III. A case study: Robust
  light curve optimization in the Large Magellanic Cloud}.
\newblock \emph{A\&A} 694, A255.
\newblock \doi{10.1051/0004-6361/202452182}
\input{das2025a}

\bibitem[{{Das} et~al.(2025{\natexlab{b}}){Das}, {Moln{\'a}r}, {Szab{\'o}},
  {Singh}, {Kanbur}, {Bhardwaj} et~al.}]{das2025b}
{Das}, S., {Moln{\'a}r}, L., {Szab{\'o}}, R., {Singh}, H.~P., {Kanbur}, S.~M.,
  {Bhardwaj}, A., et~al. (2025{\natexlab{b}}).
\newblock {A theoretical framework for BL Her stars: IV. New period-luminosity
  relations in the Rubin-LSST filters}.
\newblock \emph{A\&A} 695, A38.
\newblock \doi{10.1051/0004-6361/202452465}
\input{das2025b}

\bibitem[{{De Somma} et~al.(2024){De Somma}, {Marconi}, {Cassisi}, and
  {Molinaro}}]{desomma2024}
{De Somma}, G., {Marconi}, M., {Cassisi}, S., and {Molinaro}, R. (2024).
\newblock {Stellar Pulsation and Evolution: A Combined Theoretical Renewal and
  Updated Models (SPECTRUM). I. Updating Radiative Opacities for Pulsation
  Models of Classical Cepheid and RR-Lyrae}.
\newblock \emph{ApJ} 977, 1.
\newblock \doi{10.3847/1538-4357/ad8eb2}
\input{desomma2024}

\bibitem[{{De Somma} et~al.(2022){De Somma}, {Marconi}, {Molinaro}, {Ripepi},
  {Leccia}, and {Musella}}]{desomma2022}
{De Somma}, G., {Marconi}, M., {Molinaro}, R., {Ripepi}, V., {Leccia}, S., and
  {Musella}, I. (2022).
\newblock {An Updated Metal-dependent Theoretical Scenario for Classical
  Cepheids}.
\newblock \emph{ApJS} 262, 25.
\newblock \doi{10.3847/1538-4365/ac7f3b}
\input{desomma2022}

\bibitem[{{Deka} et~al.(2024){Deka}, {Bellinger}, {Kanbur}, {Deb}, {Bhardwaj},
  {Randall} et~al.}]{deka2024}
{Deka}, M., {Bellinger}, E.~P., {Kanbur}, S.~M., {Deb}, S., {Bhardwaj}, A.,
  {Randall}, H.~R., et~al. (2024).
\newblock {Bridging theory and observations in stellar pulsations: the impact
  of convection and metallicity on the instability strips of classical and
  type-II cepheids}.
\newblock \emph{MNRAS} 530, 5099--5119.
\newblock \doi{10.1093/mnras/stae1136}
\input{deka2024}

\bibitem[{{Demers} and {Harris}(1974)}]{demers1974}
{Demers}, S. and {Harris}, W.~E. (1974).
\newblock {The instability strip of population II cepheids.}
\newblock \emph{AJ} 79, 627--630.
\newblock \doi{10.1086/111586}
\input{demers1974}

\bibitem[{{Di Criscienzo} et~al.(2007){Di Criscienzo}, {Caputo}, {Marconi}, and
  {Cassisi}}]{dicriscienzo2007}
{Di Criscienzo}, M., {Caputo}, F., {Marconi}, M., and {Cassisi}, S. (2007).
\newblock {Synthetic properties of bright metal-poor variables. II. BL Hercules
  stars}.
\newblock \emph{A\&A} 471, 893--900.
\newblock \doi{10.1051/0004-6361:20066541}
\input{dicriscienzo2007}

\bibitem[{{Dias} et~al.(2015){Dias}, {Barbuy}, {Saviane}, {Held}, {Da Costa},
  {Ortolani} et~al.}]{dias2015}
{Dias}, B., {Barbuy}, B., {Saviane}, I., {Held}, E.~V., {Da Costa}, G.~S.,
  {Ortolani}, S., et~al. (2015).
\newblock {FORS2/VLT survey of Milky Way globular clusters. I. Description of
  the method for derivation of metal abundances in the optical and application
  to NGC 6528, NGC 6553, M 71, NGC 6558, NGC 6426, and Terzan 8}.
\newblock \emph{A\&A} 573, A13.
\newblock \doi{10.1051/0004-6361/201423996}
\input{dias2015}

\bibitem[{{Dias} et~al.(2016{\natexlab{a}}){Dias}, {Barbuy}, {Saviane}, {Held},
  {Da Costa}, {Ortolani} et~al.}]{dias2016a}
{Dias}, B., {Barbuy}, B., {Saviane}, I., {Held}, E.~V., {Da Costa}, G.~S.,
  {Ortolani}, S., et~al. (2016{\natexlab{a}}).
\newblock {FORS2/VLT survey of Milky Way globular clusters. II. Fe and Mg
  abundances of 51 Milky Way globular clusters on a homogeneous scale}.
\newblock \emph{A\&A} 590, A9.
\newblock \doi{10.1051/0004-6361/201526765}
\input{dias2016a}

\bibitem[{{Dias} et~al.(2016{\natexlab{b}}){Dias}, {Saviane}, {Barbuy}, {Held},
  {Da Costa}, {Ortolani} et~al.}]{dias2016b}
{Dias}, B., {Saviane}, I., {Barbuy}, B., {Held}, E.~V., {Da Costa}, G.,
  {Ortolani}, S., et~al. (2016{\natexlab{b}}).
\newblock {Globular Clusters and the Milky Way Connected by Chemistry}.
\newblock \emph{The Messenger} 165, 19--21
\input{dias2016b}

\bibitem[{{Ferguson} et~al.(2005){Ferguson}, {Alexander}, {Allard}, {Barman},
  {Bodnarik}, {Hauschildt} et~al.}]{ferguson2005}
{Ferguson}, J.~W., {Alexander}, D.~R., {Allard}, F., {Barman}, T., {Bodnarik},
  J.~G., {Hauschildt}, P.~H., et~al. (2005).
\newblock {Low-Temperature Opacities}.
\newblock \emph{ApJ} 623, 585--596.
\newblock \doi{10.1086/428642}
\input{ferguson2005}

\bibitem[{{Gingold}(1976)}]{gingold1976}
{Gingold}, R.~A. (1976).
\newblock {The evolutionary status of population II cepheids.}
\newblock \emph{ApJ} 204, 116--130.
\newblock \doi{10.1086/154156}
\input{gingold1976}

\bibitem[{{Gingold}(1985)}]{gingold1985}
{Gingold}, R.~A. (1985).
\newblock {The evolutionary status of Type II Cepheids}.
\newblock \emph{MmSAI} 56, 169--191
\input{gingold1985}

\bibitem[{{Gonzalez}(1994)}]{gonzalez1994}
{Gonzalez}, G. (1994).
\newblock {A Study of the UV-Bright Stars in omega CEN and the Type II Cepheid
  ST PUP}.
\newblock \emph{PASP} 106, 201.
\newblock \doi{10.1086/133370}
\input{gonzalez1994}

\bibitem[{{Gratton} et~al.(2003){Gratton}, {Bragaglia}, {Carretta},
  {Clementini}, {Desidera}, {Grundahl} et~al.}]{gratton2003}
{Gratton}, R.~G., {Bragaglia}, A., {Carretta}, E., {Clementini}, G.,
  {Desidera}, S., {Grundahl}, F., et~al. (2003).
\newblock {Distances and ages of NGC 6397, NGC 6752 and 47 Tuc}.
\newblock \emph{A\&A} 408, 529--543.
\newblock \doi{10.1051/0004-6361:20031003}
\input{gratton2003}

\bibitem[{{Green} et~al.(2019){Green}, {Schlafly}, {Zucker}, {Speagle}, and
  {Finkbeiner}}]{green2019}
{Green}, G.~M., {Schlafly}, E., {Zucker}, C., {Speagle}, J.~S., and
  {Finkbeiner}, D. (2019).
\newblock {A 3D Dust Map Based on Gaia, Pan-STARRS 1, and 2MASS}.
\newblock \emph{ApJ} 887, 93.
\newblock \doi{10.3847/1538-4357/ab5362}
\input{green2019}

\bibitem[{{Groenewegen} and {Jurkovic}(2017{\natexlab{a}})}]{groenewegen2017a}
{Groenewegen}, M.~A.~T. and {Jurkovic}, M.~I. (2017{\natexlab{a}}).
\newblock {Luminosities and infrared excess in Type II and anomalous Cepheids
  in the Large and Small Magellanic Clouds}.
\newblock \emph{A\&A} 603, A70.
\newblock \doi{10.1051/0004-6361/201730687}
\input{groenewegen2017a}

\bibitem[{{Groenewegen} and {Jurkovic}(2017{\natexlab{b}})}]{groenewegen2017}
{Groenewegen}, M.~A.~T. and {Jurkovic}, M.~I. (2017{\natexlab{b}}).
\newblock {The period-luminosity and period-radius relations of Type II and
  anomalous Cepheids in the Large and Small Magellanic Clouds}.
\newblock \emph{A\&A} 604, A29.
\newblock \doi{10.1051/0004-6361/201730946}
\input{groenewegen2017}

\bibitem[{{Harris}(1996)}]{harris1996}
{Harris}, W.~E. (1996).
\newblock {A Catalog of Parameters for Globular Clusters in the Milky Way}.
\newblock \emph{AJ} 112, 1487.
\newblock \doi{10.1086/118116}
\input{harris1996}

\bibitem[{{Hodapp} et~al.(2010){Hodapp}, {Chini}, {Reipurth}, {Murphy},
  {Lemke}, {Watermann} et~al.}]{hodapp2010}
{Hodapp}, K.~W., {Chini}, R., {Reipurth}, B., {Murphy}, M., {Lemke}, R.,
  {Watermann}, R., et~al. (2010).
\newblock {Commissioning of the infrared imaging survey (IRIS) system}.
\newblock In \emph{Ground-based and Airborne Instrumentation for Astronomy
  III}, eds. I.~S. {McLean}, S.~K. {Ramsay}, and H.~{Takami}. vol. 7735 of
  \emph{Society of Photo-Optical Instrumentation Engineers (SPIE) Conference
  Series}, 77351A.
\newblock \doi{10.1117/12.856288}
\input{hodapp2010}

\bibitem[{{Hubble} and {Humason}(1931)}]{hubble1931}
{Hubble}, E. and {Humason}, M.~L. (1931).
\newblock {The Velocity-Distance Relation among Extra-Galactic Nebulae}.
\newblock \emph{ApJ} 74, 43.
\newblock \doi{10.1086/143323}
\input{hubble1931}

\bibitem[{{Iglesias} and {Rogers}(1996)}]{iglesias1996}
{Iglesias}, C.~A. and {Rogers}, F.~J. (1996).
\newblock {Updated Opal Opacities}.
\newblock \emph{ApJ} 464, 943.
\newblock \doi{10.1086/177381}
\input{iglesias1996}

\bibitem[{{Jermyn} et~al.(2023){Jermyn}, {Bauer}, {Schwab}, {Farmer}, {Ball},
  {Bellinger} et~al.}]{jermyn2023}
{Jermyn}, A.~S., {Bauer}, E.~B., {Schwab}, J., {Farmer}, R., {Ball}, W.~H.,
  {Bellinger}, E.~P., et~al. (2023).
\newblock {Modules for Experiments in Stellar Astrophysics (MESA):
  Time-dependent Convection, Energy Conservation, Automatic Differentiation,
  and Infrastructure}.
\newblock \emph{ApJs} 265, 15.
\newblock \doi{10.3847/1538-4365/acae8d}
\input{jermyn2023}

\bibitem[{{Joy}(1949)}]{joy1949}
{Joy}, A.~H. (1949).
\newblock {Spectra of the Brighter Variables in Globular Clusters.}
\newblock \emph{ApJ} 110, 105.
\newblock \doi{10.1086/145190}
\input{joy1949}

\bibitem[{{Kiss} et~al.(2007){Kiss}, {Derekas}, {Szab{\'o}}, {Bedding}, and
  {Szabados}}]{kiss2007}
{Kiss}, L.~L., {Derekas}, A., {Szab{\'o}}, G.~M., {Bedding}, T.~R., and
  {Szabados}, L. (2007).
\newblock {Defining the instability strip of pulsating post-AGB binary stars
  from ASAS and NSVS photometry}.
\newblock \emph{MNRAS} 375, 1338--1348.
\newblock \doi{10.1111/j.1365-2966.2006.11387.x}
\input{kiss2007}

\bibitem[{{Kodric} et~al.(2018){Kodric}, {Riffeser}, {Hopp}, {Goessl}, {Seitz},
  {Bender} et~al.}]{kodric2018}
{Kodric}, M., {Riffeser}, A., {Hopp}, U., {Goessl}, C., {Seitz}, S., {Bender},
  R., et~al. (2018).
\newblock {Cepheids in M31: The PAndromeda Cepheid Sample}.
\newblock \emph{AJ} 156, 130.
\newblock \doi{10.3847/1538-3881/aad40f}
\input{kodric2018}

\bibitem[{{Kovacs} and {Buchler}(1988)}]{kovacs1988}
{Kovacs}, G. and {Buchler}, J.~R. (1988).
\newblock {Regular and Irregular Nonlinear Pulsations in Population II Cepheid
  Models}.
\newblock \emph{ApJ} 334, 971.
\newblock \doi{10.1086/166890}
\input{kovacs1988}

\bibitem[{{Kov{\'a}cs} et~al.(2023){Kov{\'a}cs}, {Nuspl}, and
  {Szab{\'o}}}]{kovacs2023}
{Kov{\'a}cs}, G.~B., {Nuspl}, J., and {Szab{\'o}}, R. (2023).
\newblock {Calibration of the convective parameters in stellar pulsation
  hydrocodes}.
\newblock \emph{MNRAS} 521, 4878--4895.
\newblock \doi{10.1093/mnras/stad884}
\input{kovacs2023}

\bibitem[{{Kov{\'a}cs} et~al.(2024){Kov{\'a}cs}, {Nuspl}, and
  {Szab{\'o}}}]{kovacs2024}
{Kov{\'a}cs}, G.~B., {Nuspl}, J., and {Szab{\'o}}, R. (2024).
\newblock {Temperature-dependent convective parameters for RRc 1D models}.
\newblock \emph{MNRAS} 527, L1--L6.
\newblock \doi{10.1093/mnrasl/slad131}
\input{kovacs2024}

\bibitem[{{Kuhfuss}(1986)}]{kuhfuss1986}
{Kuhfuss}, R. (1986).
\newblock {A model for time-dependent turbulent convection}.
\newblock \emph{A\&A} 160, 116--120
\input{kuhfuss1986}

\bibitem[{{Lejeune} et~al.(1998){Lejeune}, {Cuisinier}, and
  {Buser}}]{lejeune1998}
{Lejeune}, T., {Cuisinier}, F., and {Buser}, R. (1998).
\newblock {A standard stellar library for evolutionary synthesis. II. The M
  dwarf extension}.
\newblock \emph{A\&As} 130, 65--75.
\newblock \doi{10.1051/aas:1998405}
\input{lejeune1998}

\bibitem[{{Lengen} et~al.(2025){Lengen}, {Anderson}, {Cruz Reyes}, and
  {Viviani}}]{lengen2025}
{Lengen}, B., {Anderson}, R.~I., {Cruz Reyes}, M., and {Viviani}, G. (2025).
\newblock {A joint 1\% calibration of the RR Lyrae \& type-II Cepheid Leavitt
  laws yields homogeneous distances to 93 Galactic globular clusters}.
\newblock \emph{arXiv e-prints} ,
  arXiv:2509.16331\doi{10.48550/arXiv.2509.16331}
\input{lengen2025}

\bibitem[{{Lindegren} et~al.(2021){Lindegren}, {Klioner}, {Hern{\'a}ndez},
  {Bombrun}, {Ramos-Lerate}, {Steidelm{\"u}ller} et~al.}]{lindegren2021}
{Lindegren}, L., {Klioner}, S.~A., {Hern{\'a}ndez}, J., {Bombrun}, A.,
  {Ramos-Lerate}, M., {Steidelm{\"u}ller}, H., et~al. (2021).
\newblock {Gaia Early Data Release 3. The astrometric solution}.
\newblock \emph{A\&A} 649, A2.
\newblock \doi{10.1051/0004-6361/202039709}
\input{lindegren2021}

\bibitem[{{Maas} et~al.(2007){Maas}, {Giridhar}, and {Lambert}}]{maas2007}
{Maas}, T., {Giridhar}, S., and {Lambert}, D.~L. (2007).
\newblock {The Chemical Compositions of the Type II Cepheids-The BL Herculis
  and W Virginis Variables}.
\newblock \emph{ApJ} 666, 378--392.
\newblock \doi{10.1086/520081}
\input{maas2007}

\bibitem[{{Majaess} et~al.(2009){Majaess}, {Turner}, and {Lane}}]{majaess2009}
{Majaess}, D., {Turner}, D., and {Lane}, D. (2009).
\newblock {Type II Cepheids as Extragalactic Distance Candles}.
\newblock \emph{AcA} 59, 403--418.
\newblock \doi{10.48550/arXiv.0909.0181}
\input{majaess2009}

\bibitem[{{Majaess}(2010)}]{majaess2010}
{Majaess}, D.~J. (2010).
\newblock {RR Lyrae and Type II Cepheid Variables Adhere to a Common Distance
  Relation}.
\newblock \emph{Journal of the American Association of Variable Star Observers
  (JAAVSO)} 38, 100
\input{majaess2010}

\bibitem[{{Marconi} et~al.(2025){Marconi}, {Molinaro}, {Ripepi}, {De Somma},
  {Sicignano}, {Deka} et~al.}]{marconi2025}
{Marconi}, M., {Molinaro}, R., {Ripepi}, V., {De Somma}, G., {Sicignano}, T.,
  {Deka}, M., et~al. (2025).
\newblock {New theoretical predictions on Type II Cepheids: towards a self
  consistent Pop. II distance scale}.
\newblock \emph{arXiv e-prints} ,
  arXiv:2509.13552\doi{10.48550/arXiv.2509.13552}
\input{marconi2025}

\bibitem[{{Matsunaga} et~al.(2009){Matsunaga}, {Feast}, and
  {Menzies}}]{matsunaga2009}
{Matsunaga}, N., {Feast}, M.~W., and {Menzies}, J.~W. (2009).
\newblock {Period-luminosity relations for type II Cepheids and their
  application}.
\newblock \emph{MNRAS} 397, 933--942.
\newblock \doi{10.1111/j.1365-2966.2009.14992.x}
\input{matsunaga2009}

\bibitem[{{Matsunaga} et~al.(2011){Matsunaga}, {Feast}, and
  {Soszy{\'n}ski}}]{matsunaga2011}
{Matsunaga}, N., {Feast}, M.~W., and {Soszy{\'n}ski}, I. (2011).
\newblock {Period-luminosity relations of type II Cepheids in the Magellanic
  Clouds}.
\newblock \emph{MNRAS} 413, 223--234.
\newblock \doi{10.1111/j.1365-2966.2010.18126.x}
\input{matsunaga2011}

\bibitem[{{Matsunaga} et~al.(2006){Matsunaga}, {Fukushi}, {Nakada},
  {Tanab{\'e}}, {Feast}, {Menzies} et~al.}]{matsunaga2006}
{Matsunaga}, N., {Fukushi}, H., {Nakada}, Y., {Tanab{\'e}}, T., {Feast}, M.~W.,
  {Menzies}, J.~W., et~al. (2006).
\newblock {The period-luminosity relation for type II Cepheids in globular
  clusters}.
\newblock \emph{MNRAS} 370, 1979--1990.
\newblock \doi{10.1111/j.1365-2966.2006.10620.x}
\input{matsunaga2006}

\bibitem[{{McNamara}(1995)}]{mcnamara1995}
{McNamara}, D.~H. (1995).
\newblock {Period-Luminosity Relations of Population II Cepheids}.
\newblock \emph{AJ} 109, 2134.
\newblock \doi{10.1086/117438}
\input{mcnamara1995}

\bibitem[{{Nagashima} et~al.(1999){Nagashima}, {Nagayama}, {Nakajima},
  {Tamura}, {Sugitani}, {Nagata} et~al.}]{nagashima1999}
{Nagashima}, C., {Nagayama}, T., {Nakajima}, Y., {Tamura}, M., {Sugitani}, K.,
  {Nagata}, T., et~al. (1999).
\newblock {Development of SIRIUS --- A Simultaneous-Color InfraRed Imager for
  Unbiased Survey}.
\newblock In \emph{Star Formation 1999}, ed. T.~{Nakamoto}. 397--398
\input{nagashima1999}

\bibitem[{{Nagayama} et~al.(2003){Nagayama}, {Nagashima}, {Nakajima}, {Nagata},
  {Sato}, {Nakaya} et~al.}]{nagayama2003}
{Nagayama}, T., {Nagashima}, C., {Nakajima}, Y., {Nagata}, T., {Sato}, S.,
  {Nakaya}, H., et~al. (2003).
\newblock {SIRIUS: a near infrared simultaneous three-band camera}.
\newblock In \emph{Instrument Design and Performance for Optical/Infrared
  Ground-based Telescopes}, eds. M.~{Iye} and A.~F.~M. {Moorwood}. vol. 4841 of
  \emph{Society of Photo-Optical Instrumentation Engineers (SPIE) Conference
  Series}, 459--464.
\newblock \doi{10.1117/12.460770}
\input{nagayama2003}

\bibitem[{{Narloch} et~al.(2025){Narloch}, {Hajdu}, {Pietrzy{\'n}ski},
  {Wielg{\'o}rski}, {Smolec}, {Gieren} et~al.}]{narloch2025}
{Narloch}, W., {Hajdu}, G., {Pietrzy{\'n}ski}, G., {Wielg{\'o}rski}, P.,
  {Smolec}, R., {Gieren}, W., et~al. (2025).
\newblock {Period{\textendash}luminosity relations for Galactic Type II
  Cepheids in the Sloan bands}.
\newblock \emph{A\&A} 697, A30.
\newblock \doi{10.1051/0004-6361/202553733}
\input{narloch2025}

\bibitem[{{Nemec} et~al.(1994){Nemec}, {Nemec}, and {Lutz}}]{nemec1994}
{Nemec}, J.~M., {Nemec}, A. F.~L., and {Lutz}, T.~E. (1994).
\newblock {Period-Luminosity-Metallicity Relations, Pulsation Modes, Absolute
  Magnitudes, and Distances for Population II Variable Stars}.
\newblock \emph{AJ} 108, 222.
\newblock \doi{10.1086/117062}
\input{nemec1994}

\bibitem[{{Ngeow} et~al.(2022){Ngeow}, {Bhardwaj}, {Henderson}, {Graham},
  {Laher}, {Medford} et~al.}]{ngeow2022}
{Ngeow}, C.-C., {Bhardwaj}, A., {Henderson}, J.-Y., {Graham}, M.~J., {Laher},
  R.~R., {Medford}, M.~S., et~al. (2022).
\newblock {Zwicky Transient Facility and Globular Clusters: The
  Period-Luminosity and Period-Wesenheit Relations for Type II Cepheids}.
\newblock \emph{AJ} 164, 154.
\newblock \doi{10.3847/1538-3881/ac87a4}
\input{ngeow2022}

\bibitem[{{O'Donnell}(1994)}]{odonnell11994}
{O'Donnell}, J.~E. (1994).
\newblock {R v-dependent Optical and Near-Ultraviolet Extinction}.
\newblock \emph{ApJ} 422, 158.
\newblock \doi{10.1086/173713}
\input{odonnell11994}

\bibitem[{{Paxton} et~al.(2011){Paxton}, {Bildsten}, {Dotter}, {Herwig},
  {Lesaffre}, and {Timmes}}]{paxton2011}
{Paxton}, B., {Bildsten}, L., {Dotter}, A., {Herwig}, F., {Lesaffre}, P., and
  {Timmes}, F. (2011).
\newblock {Modules for Experiments in Stellar Astrophysics (MESA)}.
\newblock \emph{ApJS} 192, 3.
\newblock \doi{10.1088/0067-0049/192/1/3}
\input{paxton2011}

\bibitem[{{Paxton} et~al.(2013){Paxton}, {Cantiello}, {Arras}, {Bildsten},
  {Brown}, {Dotter} et~al.}]{paxton2013}
{Paxton}, B., {Cantiello}, M., {Arras}, P., {Bildsten}, L., {Brown}, E.~F.,
  {Dotter}, A., et~al. (2013).
\newblock {Modules for Experiments in Stellar Astrophysics (MESA): Planets,
  Oscillations, Rotation, and Massive Stars}.
\newblock \emph{ApJS} 208, 4.
\newblock \doi{10.1088/0067-0049/208/1/4}
\input{paxton2013}

\bibitem[{{Paxton} et~al.(2015){Paxton}, {Marchant}, {Schwab}, {Bauer},
  {Bildsten}, {Cantiello} et~al.}]{paxton2015}
{Paxton}, B., {Marchant}, P., {Schwab}, J., {Bauer}, E.~B., {Bildsten}, L.,
  {Cantiello}, M., et~al. (2015).
\newblock {Modules for Experiments in Stellar Astrophysics (MESA): Binaries,
  Pulsations, and Explosions}.
\newblock \emph{ApJS} 220, 15.
\newblock \doi{10.1088/0067-0049/220/1/15}
\input{paxton2015}

\bibitem[{{Paxton} et~al.(2018){Paxton}, {Schwab}, {Bauer}, {Bildsten},
  {Blinnikov}, {Duffell} et~al.}]{paxton2018}
{Paxton}, B., {Schwab}, J., {Bauer}, E.~B., {Bildsten}, L., {Blinnikov}, S.,
  {Duffell}, P., et~al. (2018).
\newblock {Modules for Experiments in Stellar Astrophysics (MESA): Convective
  Boundaries, Element Diffusion, and Massive Star Explosions}.
\newblock \emph{ApJS} 234, 34.
\newblock \doi{10.3847/1538-4365/aaa5a8}
\input{paxton2018}

\bibitem[{{Paxton} et~al.(2019){Paxton}, {Smolec}, {Schwab}, {Gautschy},
  {Bildsten}, {Cantiello} et~al.}]{paxton2019}
{Paxton}, B., {Smolec}, R., {Schwab}, J., {Gautschy}, A., {Bildsten}, L.,
  {Cantiello}, M., et~al. (2019).
\newblock {Modules for Experiments in Stellar Astrophysics (MESA): Pulsating
  Variable Stars, Rotation, Convective Boundaries, and Energy Conservation}.
\newblock \emph{ApJS} 243, 10.
\newblock \doi{10.3847/1538-4365/ab2241}
\input{paxton2019}

\bibitem[{{Planck Collaboration} et~al.(2020){Planck Collaboration}, {Aghanim,
  N.}, {Akrami, Y.}, {Ashdown, M.}, {Aumont, J.}, {Baccigalupi, C.}
  et~al.}]{planck2020}
{Planck Collaboration}, {Aghanim, N.}, {Akrami, Y.}, {Ashdown, M.}, {Aumont,
  J.}, {Baccigalupi, C.}, et~al. (2020).
\newblock Planck 2018 results - vi. cosmological parameters.
\newblock \emph{A\&A} 641, A6.
\newblock \doi{10.1051/0004-6361/201833910}
\input{planck2020}

\bibitem[{{Pritzl} et~al.(2002){Pritzl}, {Smith}, {Catelan}, and
  {Sweigart}}]{pritzl2002}
{Pritzl}, B.~J., {Smith}, H.~A., {Catelan}, M., and {Sweigart}, A.~V. (2002).
\newblock {Variable Stars in the Unusual, Metal-rich Globular Cluster NGC
  6388}.
\newblock \emph{AJ} 124, 949--976.
\newblock \doi{10.1086/341381}
\input{pritzl2002}

\bibitem[{{Ramolla} et~al.(2016){Ramolla}, {Westhues}, {Hackstein}, {Haas},
  {Hodapp}, {Lemke} et~al.}]{ramolla2016}
{Ramolla}, M., {Westhues}, C., {Hackstein}, M., {Haas}, M., {Hodapp}, K.,
  {Lemke}, R., et~al. (2016).
\newblock {A green observatory in the Chilean Atacama desert}.
\newblock In \emph{Modeling, Systems Engineering, and Project Management for
  Astronomy VI}, eds. G.~Z. {Angeli} and P.~{Dierickx}. vol. 9911 of
  \emph{Society of Photo-Optical Instrumentation Engineers (SPIE) Conference
  Series}, 99112M.
\newblock \doi{10.1117/12.2234018}
\input{ramolla2016}

\bibitem[{{Riess} et~al.(2025){Riess}, {Li}, {Anand}, {Yuan}, {Breuval},
  {Casertano} et~al.}]{riess2025}
{Riess}, A.~G., {Li}, S., {Anand}, G.~S., {Yuan}, W., {Breuval}, L.,
  {Casertano}, S., et~al. (2025).
\newblock {The Perfect Host: JWST Cepheid Observations in a Background-Free SN
  Ia Host Confirm No Bias in Hubble-Constant Measurements}.
\newblock \emph{arXiv e-prints} ,
  arXiv:2509.01667\doi{10.48550/arXiv.2509.01667}
\input{riess2025}

\bibitem[{{Riess} et~al.(2009){Riess}, {Macri}, {Casertano}, {Sosey},
  {Lampeitl}, {Ferguson} et~al.}]{riess2009}
{Riess}, A.~G., {Macri}, L., {Casertano}, S., {Sosey}, M., {Lampeitl}, H.,
  {Ferguson}, H.~C., et~al. (2009).
\newblock {A Redetermination of the Hubble Constant with the Hubble Space
  Telescope from a Differential Distance Ladder}.
\newblock \emph{ApJ} 699, 539--563.
\newblock \doi{10.1088/0004-637X/699/1/539}
\input{riess2009}

\bibitem[{{Riess} et~al.(2024){Riess}, {Scolnic}, {Anand}, {Breuval},
  {Casertano}, {Macri} et~al.}]{riess2024}
{Riess}, A.~G., {Scolnic}, D., {Anand}, G.~S., {Breuval}, L., {Casertano}, S.,
  {Macri}, L.~M., et~al. (2024).
\newblock {JWST Validates HST Distance Measurements: Selection of Supernova
  Subsample Explains Differences in JWST Estimates of Local H $_{0}$}.
\newblock \emph{ApJ} 977, 120.
\newblock \doi{10.3847/1538-4357/ad8c21}
\input{riess2024}

\bibitem[{{Riess} et~al.(2022){Riess}, {Yuan}, {Macri}, {Scolnic}, {Brout},
  {Casertano} et~al.}]{riess2022}
{Riess}, A.~G., {Yuan}, W., {Macri}, L.~M., {Scolnic}, D., {Brout}, D.,
  {Casertano}, S., et~al. (2022).
\newblock {A Comprehensive Measurement of the Local Value of the Hubble
  Constant with 1 km s$^{-1}$ Mpc$^{-1}$ Uncertainty from the Hubble Space
  Telescope and the SH0ES Team}.
\newblock \emph{ApJl} 934, L7.
\newblock \doi{10.3847/2041-8213/ac5c5b}
\input{riess2022}

\bibitem[{{Ripepi} et~al.(2023){Ripepi}, {Clementini}, {Molinaro}, {Leccia},
  {Plachy}, {Moln{\'a}r} et~al.}]{ripepi2023}
{Ripepi}, V., {Clementini}, G., {Molinaro}, R., {Leccia}, S., {Plachy}, E.,
  {Moln{\'a}r}, L., et~al. (2023).
\newblock {Gaia Data Release 3. Specific processing and validation of all sky
  RR Lyrae and Cepheid stars: The Cepheid sample}.
\newblock \emph{A\&A} 674, A17.
\newblock \doi{10.1051/0004-6361/202243990}
\input{ripepi2023}

\bibitem[{{Ripepi} et~al.(2025){Ripepi}, {Trentin}, {Catanzaro}, {Marconi},
  {Bhardwaj}, {Clementini} et~al.}]{ripepi2025}
{Ripepi}, V., {Trentin}, E., {Catanzaro}, G., {Marconi}, M., {Bhardwaj}, A.,
  {Clementini}, G., et~al. (2025).
\newblock {Cepheid Metallicity in the Leavitt Law (C--MetaLL) survey: VII.
  Metallicity dependence of Period-Wesenheit relations based on a homogeneous
  spectroscopic sample}.
\newblock \emph{arXiv e-prints} ,
  arXiv:2508.17447\doi{10.48550/arXiv.2508.17447}
\input{ripepi2025}

\bibitem[{{Rodrigo} and {Solano}(2020)}]{rodrigo2020}
{Rodrigo}, C. and {Solano}, E. (2020).
\newblock {The SVO Filter Profile Service}.
\newblock In \emph{XIV.0 Scientific Meeting (virtual) of the Spanish
  Astronomical Society}. 182
\input{rodrigo2020}

\bibitem[{{Rodrigo} et~al.(2012){Rodrigo}, {Solano}, and {Bayo}}]{rodrigo2012}
[Dataset] {Rodrigo}, C., {Solano}, E., and {Bayo}, A. (2012).
\newblock {SVO Filter Profile Service Version 1.0}.
\newblock IVOA Working Draft 15 October 2012.
\newblock \doi{10.5479/ADS/bib/2012ivoa.rept.1015R}
\input{rodrigo2012}

\bibitem[{{Sandage} and {Tammann}(2006)}]{sandage2006}
{Sandage}, A. and {Tammann}, G.~A. (2006).
\newblock {Absolute Magnitude Calibrations of Population I and II Cepheids and
  Other Pulsating Variables in the Instability Strip of the Hertzsprung-Russell
  Diagram}.
\newblock \emph{ARA\&A} 44, 93--140.
\newblock \doi{10.1146/annurev.astro.43.072103.150612}
\input{sandage2006}

\bibitem[{{Schlafly} and {Finkbeiner}(2011)}]{schlafly2011}
{Schlafly}, E.~F. and {Finkbeiner}, D.~P. (2011).
\newblock {Measuring Reddening with Sloan Digital Sky Survey Stellar Spectra
  and Recalibrating SFD}.
\newblock \emph{ApJ} 737, 103.
\newblock \doi{10.1088/0004-637X/737/2/103}
\input{schlafly2011}

\bibitem[{{Schlegel} et~al.(1998){Schlegel}, {Finkbeiner}, and
  {Davis}}]{schlegel1998}
{Schlegel}, D.~J., {Finkbeiner}, D.~P., and {Davis}, M. (1998).
\newblock {Maps of Dust Infrared Emission for Use in Estimation of Reddening
  and Cosmic Microwave Background Radiation Foregrounds}.
\newblock \emph{ApJ} 500, 525--553.
\newblock \doi{10.1086/305772}
\input{schlegel1998}

\bibitem[{{Sicignano} et~al.(2024){Sicignano}, {Ripepi}, {Marconi}, {Molinaro},
  {Bhardwaj}, {Cioni} et~al.}]{sicignano2024}
{Sicignano}, T., {Ripepi}, V., {Marconi}, M., {Molinaro}, R., {Bhardwaj}, A.,
  {Cioni}, M. R.~L., et~al. (2024).
\newblock {The VMC survey. L. Type II Cepheids in the Magellanic Clouds:
  Period-luminosity relations in the near-infrared bands}.
\newblock \emph{A\&A} 685, A41.
\newblock \doi{10.1051/0004-6361/202348650}
\input{sicignano2024}

\bibitem[{{Smolec}(2016)}]{smolec2016}
{Smolec}, R. (2016).
\newblock {Survey of non-linear hydrodynamic models of type-II Cepheids}.
\newblock \emph{MNRAS} 456, 3475--3493.
\newblock \doi{10.1093/mnras/stv2868}
\input{smolec2016}

\bibitem[{{Smolec} and {Moskalik}(2008)}]{smolec2008}
{Smolec}, R. and {Moskalik}, P. (2008).
\newblock {Convective Hydrocodes for Radial Stellar Pulsation. Physical and
  Numerical Formulation}.
\newblock \emph{AcA} 58, 193--232
\input{smolec2008}

\bibitem[{{Soszy{\'n}ski} et~al.(2019){Soszy{\'n}ski}, {Smolec}, {Udalski}, and
  {Pietrukowicz}}]{soszynski2019}
{Soszy{\'n}ski}, I., {Smolec}, R., {Udalski}, A., and {Pietrukowicz}, P.
  (2019).
\newblock {Type II Cepheids Pulsating in the First Overtone from the OGLE
  Survey}.
\newblock \emph{ApJ} 873, 43.
\newblock \doi{10.3847/1538-4357/ab04ab}
\input{soszynski2019}

\bibitem[{{Soszy{\'n}ski} et~al.(2011){Soszy{\'n}ski}, {Udalski},
  {Pietrukowicz}, {Szyma{\'n}ski}, {Kubiak}, {Pietrzy{\'n}ski}
  et~al.}]{soszynski2011}
{Soszy{\'n}ski}, I., {Udalski}, A., {Pietrukowicz}, P., {Szyma{\'n}ski}, M.~K.,
  {Kubiak}, M., {Pietrzy{\'n}ski}, G., et~al. (2011).
\newblock {The Optical Gravitational Lensing Experiment. The OGLE-III Catalog
  of Variable Stars. XIV. Classical and TypeII Cepheids in the Galactic Bulge}.
\newblock \emph{AcA} 61, 285--301
\input{soszynski2011}

\bibitem[{{Soszy{\'n}ski} et~al.(2008){Soszy{\'n}ski}, {Udalski},
  {Szyma{\'n}ski}, {Kubiak}, {Pietrzy{\'n}ski}, {Wyrzykowski}
  et~al.}]{soszynski2008}
{Soszy{\'n}ski}, I., {Udalski}, A., {Szyma{\'n}ski}, M.~K., {Kubiak}, M.,
  {Pietrzy{\'n}ski}, G., {Wyrzykowski}, {\L}., et~al. (2008).
\newblock {The Optical Gravitational Lensing Experiment. The OGLE-III Catalog
  of Variable Stars. II.Type II Cepheids and Anomalous Cepheids in the Large
  Magellanic Cloud}.
\newblock \emph{AcA} 58, 293
\input{soszynski2008}

\bibitem[{{Soszy{\'n}ski} et~al.(2014){Soszy{\'n}ski}, {Udalski},
  {Szyma{\'n}ski}, {Pietrukowicz}, {Mr{\'o}z}, {Skowron}
  et~al.}]{soszynski2014}
{Soszy{\'n}ski}, I., {Udalski}, A., {Szyma{\'n}ski}, M.~K., {Pietrukowicz}, P.,
  {Mr{\'o}z}, P., {Skowron}, J., et~al. (2014).
\newblock {Over 38000 RR Lyrae Stars in the OGLE Galactic Bulge Fields}.
\newblock \emph{AcA} 64, 177--196
\input{soszynski2014}

\bibitem[{{Soszy{\'n}ski} et~al.(2017){Soszy{\'n}ski}, {Udalski},
  {Szyma{\'n}ski}, {Wyrzykowski}, {Ulaczyk}, {Poleski} et~al.}]{soszynski2017}
{Soszy{\'n}ski}, I., {Udalski}, A., {Szyma{\'n}ski}, M.~K., {Wyrzykowski},
  {\L}., {Ulaczyk}, K., {Poleski}, R., et~al. (2017).
\newblock {The OGLE Collection of Variable Stars. Classical, Type II, and
  Anomalous Cepheids toward the Galactic Center}.
\newblock \emph{AcA} 67, 297--316
\input{soszynski2017}

\bibitem[{{Soszy{\'n}ski} et~al.(2018){Soszy{\'n}ski}, {Udalski},
  {Szyma{\'n}ski}, {Wyrzykowski}, {Ulaczyk}, {Poleski} et~al.}]{soszynski2018}
{Soszy{\'n}ski}, I., {Udalski}, A., {Szyma{\'n}ski}, M.~K., {Wyrzykowski},
  {\L}., {Ulaczyk}, K., {Poleski}, R., et~al. (2018).
\newblock {The OGLE Collection of Variable Stars. Type II Cepheids in the
  Magellanic System}.
\newblock \emph{AcA} 68, 89--109
\input{soszynski2018}

\bibitem[{{Stellingwerf}(1982{\natexlab{a}})}]{stellingwerf1982a}
{Stellingwerf}, R.~F. (1982{\natexlab{a}}).
\newblock {Convection in pulsating stars. I. Non linear hydro-dynamics.}
\newblock \emph{ApJ} 262, 330--338.
\newblock \doi{10.1086/160425}
\input{stellingwerf1982a}

\bibitem[{{Stellingwerf}(1982{\natexlab{b}})}]{stellingwerf1982b}
{Stellingwerf}, R.~F. (1982{\natexlab{b}}).
\newblock {Convection in pulsating stars. II. RR LYR convection and stability.}
\newblock \emph{ApJ} 262, 339--343.
\newblock \doi{10.1086/160426}
\input{stellingwerf1982b}

\bibitem[{{V{\'a}squez} et~al.(2018){V{\'a}squez}, {Saviane}, {Held}, {Da
  Costa}, {Dias}, {Gullieuszik} et~al.}]{vasquez2018}
{V{\'a}squez}, S., {Saviane}, I., {Held}, E.~V., {Da Costa}, G.~S., {Dias}, B.,
  {Gullieuszik}, M., et~al. (2018).
\newblock {Homogeneous metallicities and radial velocities for Galactic
  globular clusters. II. New CaT metallicities for 28 distant and reddened
  globular clusters}.
\newblock \emph{A\&A} 619, A13.
\newblock \doi{10.1051/0004-6361/201833525}
\input{vasquez2018}

\bibitem[{{Verde} et~al.(2019){Verde}, {Treu}, and {Riess}}]{verde2019}
{Verde}, L., {Treu}, T., and {Riess}, A.~G. (2019).
\newblock {Tensions between the early and late Universe}.
\newblock \emph{Nature Astronomy} 3, 891--895.
\newblock \doi{10.1038/s41550-019-0902-0}
\input{verde2019}

\bibitem[{{Vinko} et~al.(1998){Vinko}, {Remage Evans}, {Kiss}, and
  {Szabados}}]{vinko1998}
{Vinko}, J., {Remage Evans}, N., {Kiss}, L.~L., and {Szabados}, L. (1998).
\newblock {Spectroscopic survey of field Type II Cepheids}.
\newblock \emph{MNRAS} 296, 824--838.
\newblock \doi{10.1046/j.1365-8711.1998.01389.x}
\input{vinko1998}

\bibitem[{{Wallerstein}(2002)}]{wallerstein2002}
{Wallerstein}, G. (2002).
\newblock {The Cepheids of Population II and Related Stars}.
\newblock \emph{PASP} 114, 689--699.
\newblock \doi{10.1086/341698}
\input{wallerstein2002}

\bibitem[{{Wallerstein} and {Cox}(1984)}]{wallerstein1984}
{Wallerstein}, G. and {Cox}, A.~N. (1984).
\newblock {The population II Cepheids.}
\newblock \emph{PASP} 96, 677--691.
\newblock \doi{10.1086/131406}
\input{wallerstein1984}

\bibitem[{{Wielg{\'o}rski} et~al.(2024){Wielg{\'o}rski}, {Pietrzy{\'n}ski},
  {Gieren}, {Zgirski}, {G{\'o}rski}, {Storm} et~al.}]{wielgorski2024}
{Wielg{\'o}rski}, P., {Pietrzy{\'n}ski}, G., {Gieren}, W., {Zgirski}, B.,
  {G{\'o}rski}, M., {Storm}, J., et~al. (2024).
\newblock {Projection factor and radii of Type II Cepheids: BL Her stars}.
\newblock \emph{A\&A} 689, A241.
\newblock \doi{10.1051/0004-6361/202450182}
\input{wielgorski2024}

\bibitem[{{Wielg{\'o}rski} et~al.(2022){Wielg{\'o}rski}, {Pietrzy{\'n}ski},
  {Pilecki}, {Gieren}, {Zgirski}, {G{\'o}rski} et~al.}]{wielgorski2022}
{Wielg{\'o}rski}, P., {Pietrzy{\'n}ski}, G., {Pilecki}, B., {Gieren}, W.,
  {Zgirski}, B., {G{\'o}rski}, M., et~al. (2022).
\newblock {An Absolute Calibration of the Near-infrared Period-Luminosity
  Relations of Type II Cepheids in the Milky Way and in the Large Magellanic
  Cloud}.
\newblock \emph{ApJ} 927, 89.
\newblock \doi{10.3847/1538-4357/ac470c}
\input{wielgorski2022}

\bibitem[{{Yacob} et~al.(2025){Yacob}, {Berdinkov}, and
  {Pastukhova}}]{yacob2025}
{Yacob}, A.~M., {Berdinkov}, L.~N., and {Pastukhova}, E.~N. (2025).
\newblock {A study on the evolutionary period changes of short-period type II
  cepheids}.
\newblock \emph{Ap\&SS} 370, 28.
\newblock \doi{10.1007/s10509-025-04418-7}
\input{yacob2025}

\bibitem[{{Zinn} and {West}(1984)}]{zinn1984}
{Zinn}, R. and {West}, M.~J. (1984).
\newblock {The globular cluster system of the Galaxy. III. Measurements of
  radial velocity and metallicity for 60 clusters and a compilation of
  metallicities for 121 clusters.}
\newblock \emph{ApJs} 55, 45--66.
\newblock \doi{10.1086/190947}
\input{zinn1984}

\end{thebibliography}
\end{document}